\documentclass[preprint, superscriptaddress,amsmath, nofootinbib]{revtex4-1}
\usepackage{graphicx}
\usepackage{latexsym}
\usepackage{slashed}
\usepackage{bm}
\usepackage{color}
\usepackage[colorlinks,citecolor=blue,urlcolor=blue,linkcolor=blue]{hyperref}
\usepackage{diagbox}

\def\({\left(}
\def\){\right)}

\def\beq{\begin{equation}}
\def\eeq{\end{equation}}

\begin{document}

\title{Axion Dark Radiation: Hubble Tension and Hyper-kamiokande Neutrino Experiment}
	
\author{Yuchao Gu}
\email{guyc@njnu.edu.cn}
\affiliation{Department of Physics and Institute of Theoretical Physics, Nanjing Normal University, Nanjing, 210023, China}

\author{Lei Wu}
\email{leiwu@njnu.edu.cn}
\affiliation{Department of Physics and Institute of Theoretical Physics, Nanjing Normal University, Nanjing, 210023, China}

\author{Bin Zhu}
\email{zhubin@mail.nankai.edu.cn}
\affiliation{Department of Physics, Yantai University, Yantai 264005, China}

\begin{abstract}
In this work, we investigate the dark sector of a supersymmetric axion model, consisting of the late-decaying gravitino/axino dark matter and axion dark radiation. In the early universe, the decay of the scalar superpartner of the axion (saxion) will produce a large amount of entropy. The additional entropy can not only dilute the relic density of the gravitino/axino dark matter to avoid overclosing the universe but also relax the constraint on the reheating temperature $T_{R}$ after inflation. Meanwhile, the axion dark radiation from the saxion decay will increase the effective number of neutrino species $N_{\rm eff}$, which can help to reduce the cosmological Hubble tension. In the late universe, the decay of long-lived gravitino/axino dark matter produces the axions with MeV-scale kinetic energy. We study the potential of searching for such energetic axions through the inverse Primakoff process $a+A \to \gamma+ A$ in the neutrino experiments, such as Hyper-Kamiokande.
\end{abstract}

\maketitle

\tableofcontents

\newpage
\section{Introduction}
The Standard Model (SM) of particle physics is a successful theory that describes the currently known elemental particles and their interactions. Despite its great success, there are still several problems in the Standard Model, such as the absence of Dark Matter (DM) candidates and strong CP problem~\cite{Peccei:1977hh}. Supersymmetry is one of the most promising extensions of the SM that can address these fundamental problems. When $R$-parity is conserved, the lightest supersymmetric particle (LSP) can play the role of DM particle naturally. In addition, the supersymmetric axion models based on the Peccei-Quinn mechanism also provide an elegant solution to the strong CP problem~\cite{Kim:1979if,Shifman:1979if,Zhitnitsky:1980tq,Dine:1981rt,Kim:1983dt,Kim:2008hd}.

The lightest neutralino as Weakly Interacting Massive Dark Matter (WIMP) has been extensively studied in the past decades. However, the null results of searching for WIMP dark matter have put stringent constraints on its mass and interaction strength~\cite{Cao:2013mqa, Duan:2017ucw, Abdughani:2017dqs, Duan:2018rls, Abdughani:2019wss, Abdughani:2019wai, Abdughani:2021pdc}, which strongly motivates the theoretical and experimental community to search for light or super-weakly interacting DM particles. In the supersymmetric models, the gravitino (the superpartner of graviton) or axino (the superpartner of axion) can serve as such a kind of feeble DM particles (super-WIMP)~\cite{Covi:1999ty,Covi:2001nw,Covi:2009pq,Bae:2014efa,Co:2016fln,Hamaguchi:2017ihw}. Depending on the SUSY breaking scheme, the masses of these DM particles could be tiny. Their couplings with the SM particles are extremely weak because of the large suppression of Planck scale $M_{P}$ or the PQ symmetry breaking scale $V_{\rm PQ}$.

More interestingly, these super-WIMPs may have a connection with some tentative cosmological anomalies, such as the Hubble tension~\cite{Hooper:2011aj,Gu:2020ozv,Anchordoqui:2020djl}, which refers to the tension that the value of Hubble constant $H_{0}=(67.36 \pm 0.54)$ km s$^{-1}$Mpc$^{-1}$~\cite{Verde:2019ivm} inferred from the cosmic microwave background (CMB) in the early universe shows 4$\sigma$ to 6$\sigma$ discrepancy from that derived from the local distance ladder in the late universe, $H_{0}=(74.03 \pm 1.42)$ km s$^{-1}$ Mpc$^{-1}$~\cite{Riess:2019cxk}. Many attractive ideas~\cite{DiValentino:2021izs}, such as the early dark energy~\cite{Karwal:2016vyq,Gogoi:2020qif}  and the late decaying dark matter~\cite{Ichiki:2004vi,Bjaelde:2012wi}, have been proposed to solve the Hubble constant problem. However, when other cosmological observables are accounted for, all existing proposals can only help to alleviate this tension to some extent~\cite{Riess:2019cxk,Riess:2021jrx}. Among them, dark radiation~\cite{Binder:2017lkj, Pandey:2019plg, Vattis:2019efj, Xiao:2019ccl, Blinov:2020uvz,Dainotti:2021pqg} is one of the possible ways to reduce the tension, although a perfect solution is still under investigation. The extra relativistic degrees of freedom can increase the effective number of neutrino species $N_{\rm eff}$ and thus reduce the sound horizon $r_s(z_{LS})$ during the time up to the last scattering. While the angular scale $\theta_\star=r_s(z_{LS})/D_A(z_{LS})$ is fixed by the measured acoustic peaks from the CMB power spectrum, where $D_A$ is the angular diameter. Therefore, decreasing $D_A$ can keep $\theta_\star$ unchanged, which can be achieved by simultaneously increasing $H_0$.  

In this paper, we investigate the phenomenology of the dark sector in a supersymmetric axion model, where the gravitino/axino plays the role of late-decaying DM. In general, there will exist a serious cosmological gravitino/axino problem of overclosure via thermal production~\cite{Pagels:1981ke,Moroi:1993mb,Asaka:2000zh,Bolz:2000fu,Roszkowski:2004jd,Pradler:2006hh,Pradler:2006qh,Cheung:2011mg,Co:2017orl,Gu:2020ozv}. Fortunately, the supersymmetry breaking provides the saxion field with a potential after inflation, which results in the saxion dominating the early universe. With the expansion of the universe, the saxion begins to decay during the dominant epoch. Saxion decay ends at the temperature $T_{s}$, followed by the radiation dominated era. But thanks to the saxion decay, a large amount of entropy can be produced to dilute the relic density of the gravitino/axino dark matter in our model. Besides, the axions from the saxion decay in the early universe can serve as dark radiation~\cite{Choi:1996vz,Hashimoto:1998ua,Chun:2000jr,Ichikawa:2007jv}, which can increase the effective neutrino species and thus relax the Hubble tension. However, it is mentioned that too much dark radiation produced in the early universe may affect predictions of the CMB power spectrum and big bang nucleosynthesis(BBN) since extra dark radiation will accelerate the expansion rate of the universe. Furthermore, if the late-decaying DM decays to the LSP plus the SM particle directly, it may result in the non-thermal nuclear reactions during or after BBN~\cite{Kawasaki:2004qu}, which can also destroy the perfect predictions of light elements abundance. But in our scenario, if the gravitino (axino) is the NLSP, it will decay into axino (gravitino) LSP and axion so that the strong constraints from BBN can be escaped. More interestingly, the axion from the decay of gravitino/axino in the late universe can be tested through the absorptive effects in the new direct detection experiments~\cite{Hochberg:2016sqx,Cui:2017ytb,Kuo:2021mtp} or through the inverse Primakoff scattering~\cite{Avignone:1997th,Bernabei:2001ny,Dent:2020jhf} in neutrino experiments, such as Hyper-Kamiokande~\cite{Abe:2018uyc}, DUNE~\cite{Acciarri:2015uup,Abi:2018dnh,Abi:2020kei} and JUNO~\cite{An:2015jdp}. 

This paper is laid out as follows. In Section~\ref{sec2}, we briefly introduce our SUSY DFSZ axion model and relevant interactions. In Section~\ref{sec3}, we discuss the saxion decay and implication for the Hubble tension. In Section~\ref{sec4}, we will consider the dilution effect originating from saxion decay and calculate the relic density of gravitino/axino DM. In section~\ref{sec5}, we will investigate the possibility of detecting the axion with MeV kinetic energy from the late decaying gravitino DM through the inverse Primakoff
scattering in neutrino experiments. 
Finally, we draw some conclusions in Section~\ref{sec6}.

\section{SUSY DFSZ axion model}
\label{sec2}

In the supersymmetric axion model, the axion superfield at the energy scale below the PQ symmetry breaking scale $V_{PQ}$ is given by
\begin{equation}
     \hat{A}=\frac{s+ia}{\sqrt{2}}+\sqrt{2}\theta\tilde{a}+\theta^{2}F,
\end{equation}
where $s$ is the saxion field, $a$ is the axion field and $\tilde{a}$ is the fermionic superpartner of axion i.e. axino. The advantage of the SUSY DFSZ model is that the $\mu$ problem and the strong CP problem can be solved simultaneously by the well-known Kim-Nilles mechanism~\cite{Kim:1983dt}. Since two Higgs doublets carry the PQ charge, the bare $\mu$ term is forbidden. Only the higher dimensional operator is allowed,
\begin{equation}
    W_{\rm DFSZ} \ni \lambda \frac{{\hat{S}}^2}{M_{P}} \hat{H}_{u} \hat{H}_{d}=\mu \hat{H}_u\hat{H}_d.
\end{equation}
When the PQ symmetry is broken, the singlet superfield acquires the vacuum expectation value $\langle S \rangle \sim V_{PQ}$. Then the higgsino can obtain the mass $\mu \sim \lambda V_{PQ}^2/M_{pl}$. Interestingly, a natural $\mu$ term with $\mathcal{O}(10^2)$ GeV can be obtained by assuming $V_{PQ}\sim \mathcal{O}(10^{10})$ GeV and $\lambda \sim \mathcal{O}(1)$. 
Therefore, the interaction of the axion superfield and the Higgs fields can be expressed by the spurion analysis from the following superpotential,
\begin{equation}
    \mathcal{L}_{\rm DFSZ}=\int d^2\theta\, \mu \exp\bigg[c_{\mu}\frac{\hat{A}}{V_{PQ}}\bigg] \hat{H}_{u} \hat{H}_{d}\sim \int d^2\theta \mu \hat{H}_u \hat{H}_d + \mu\frac{c_{\mu}\hat A}{V_{PQ}} \hat{H}_u \hat{H}_d+ \cdots.
    \label{eq:DFSZ}
\end{equation}
Here $c_{\mu}$ is the PQ charge of Higgs doublets bi-linear operator, which is usually taken as $c_\mu=2$ because the Higgs field is charged by $+1$. In addition, the interactions of saxion-axion-axino and the saxion self-interaction come from the K\"{a}hler potential 
\begin{equation}
    K=\sum_{i} v_{i}^2 {\rm exp}\bigg[c_{i}\left(\frac{\hat{A}+\hat{A}^{\dagger}}{V_{PQ}}\right)\bigg]
\label{kahler}
\end{equation} 
By expanding Eq.~\ref{kahler} up to the cubic term, we can obtain the relevant saxion interactions in our study,
\begin{equation}
    \mathcal{L}_{saa,s \tilde{a}\tilde{a}}=-\frac{\kappa}{\sqrt{2} V_{PQ}}s \partial_{\mu}a \partial^{\mu}a-\frac{\kappa}{\sqrt{2} V_{PQ}}s \partial_{\mu}s \partial^{\mu}s+\frac{\kappa}{\sqrt{2} V_{PQ}}s m_{\tilde{a}}\left(\tilde{a} \tilde{a}+\tilde{a}^{\dagger}\tilde{a}^{\dagger}\right),
\end{equation}
where $\kappa=\sum_{i}c_{i}^3 v_{i}^2/V_{PQ}^2$ is dimensionless coupling. We assume a single PQ charge and take $\kappa=1$ in our calculation.

\begin{figure}[h]
\centering
\includegraphics[height=7cm,width=7cm]{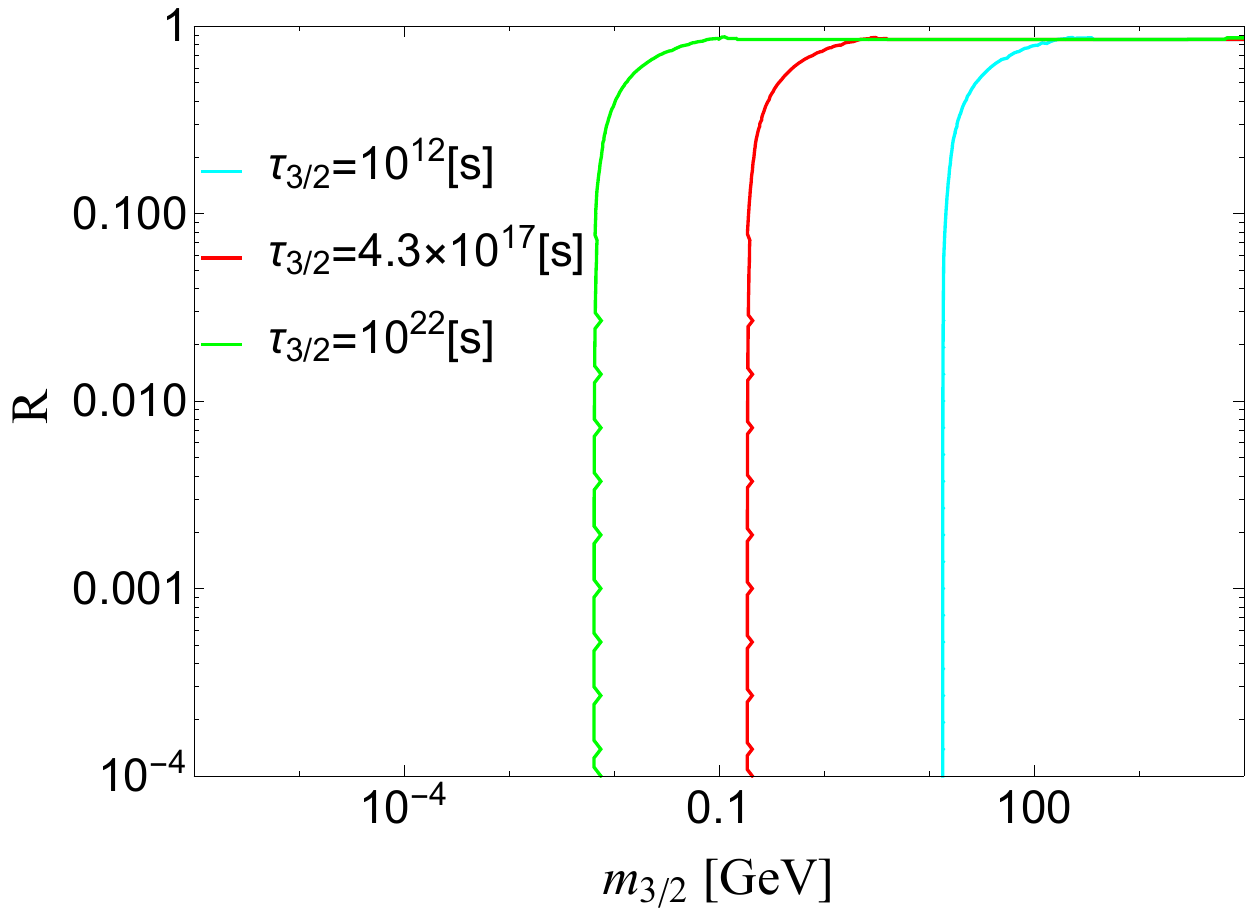}
\hspace{0.5cm}
\includegraphics[height=7cm,width=7cm]{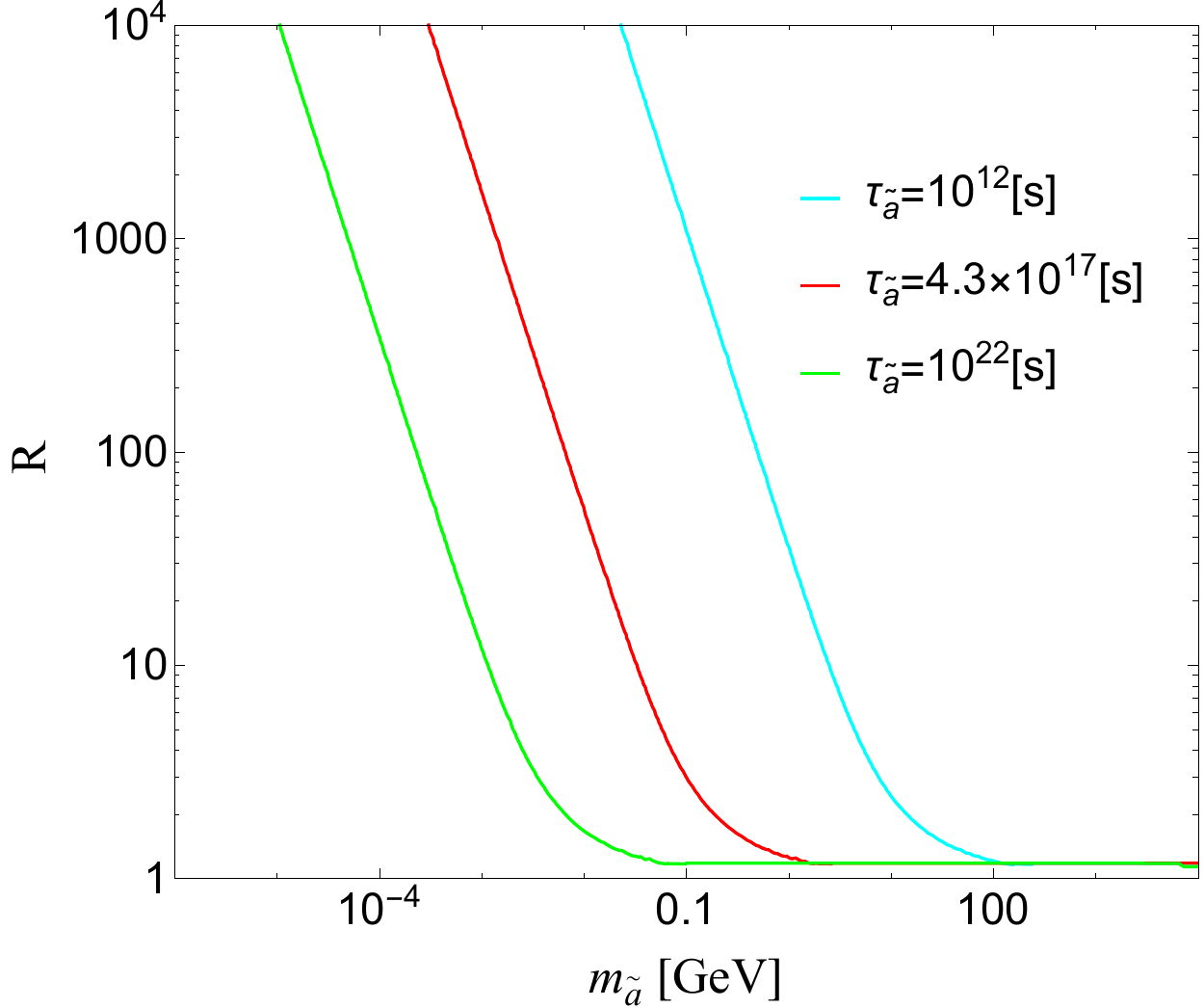}
\caption{The lifetimes of the gravitino/axino on the plane of mass ratio $R$ versus the gravitino mass $m_{3/2}$ and the axino mass $m_{\tilde{a}}$, respectively.}
\label{lifetime}
\end{figure}

Besides, the gravitino exists naturally in SUSY axion model, which is the prediction of the local supersymmetry. Its interactions with the SM particles is given by,
\begin{equation}
    \mathcal{L}_{\tilde{G}}=-\frac{i}{2 M_{P}} \mathcal{J}^{\mu} \tilde{G}+h.c.
\end{equation}
with
\begin{equation}
    \mathcal{J}^{\mu}=\sqrt{2} \sigma^{\nu} \Bar{\sigma}^{\mu}\chi \partial_{\nu} \phi^{\dagger}-\sigma^{\alpha \beta}\sigma^{\mu}\lambda^{a \dagger} F_{\alpha \beta}^a
\end{equation}
where the supercurrent $\mathcal{J^{\mu}}$ includes the gauge supermultiplets $(V,\lambda)$ and the chiral supermultiplets $(\phi,\chi)$. This will induce the interaction of the gravitino with the axion supermultiplet. Then, we can derive the interaction of the gravitino, the axino and the axion,
\begin{equation}
    \mathcal{L}=-\frac{1}{2 M_{P}} \partial_{\nu}a \Bar{\psi}_{\mu} \gamma^{\nu} \gamma^{\mu}i \gamma_{5}\tilde{a}
\end{equation}
where the $\psi_{\mu}$ is the gravitino field. If the gravitino is heavier than the axino, the decay channel $\tilde{G} \rightarrow a \tilde{a}$ is allowed and the corresponding decay width is given by
\begin{equation}
    \Gamma_{3/2}=\frac{m_{3/2}^3}{192 \pi M_{P}^2}(1-R)^2(1-R^2)^3
\end{equation}
where $R\equiv m_{\tilde{a}}/m_{3/2}$ is the mass ratio. On the contrary, the axino will decay to the gravitino via the process $\tilde{a} \to \tilde{G}a$ provided that $m_{\tilde{a}}$ is larger than $m_{3/2}$. In this case, the decay width of the axino is given by
\begin{equation}
    \Gamma_{\tilde{a}}=\frac{m_{\tilde{a}}^5}{96 \pi m_{3/2}^2 M_{P}^2}\left(1-R^{-1}\right)^2 \left(1-R^{-2}\right)^3
\end{equation}

In Fig.~\ref{lifetime}, we show the numerical results of the lifetimes of the gravitino ($\tau_{3/2}$) and the axino ($\tau_{\tilde{a}}$) for the above two cases on the plane of the mass ratio $R$ versus the gravitino mass $m_{3/2}$ and axino mass $m_{\tilde{a}}$ respectively. The green, red and cyan lines illustrate the lifetime $\tau=10^{12}$, $4.3 \times 10^{17}$ and $10^{22}$s, respectively. It can be seen that the lifetime of the gravitino/axino will increase with the decrease of its mass for a given value of the mass ratio $R$. When the gravitino/axino mass approaches to the LSP mass, i.e. $R \rightarrow 1$, their lifetimes will be independent of their masses. On the other hand, when $R \rightarrow \infty$, the lifetime of the axino depends not only on its mass $m_{\tilde{a}}$ but also the mass ratio $R$. While if $R \rightarrow 0$, the lifetime of the gravitino only depends on its mass $m_{3/2}$. These features will lead to the restriction on our axion from the decay of these late-decaying DMs.

\begin{figure}[ht]
\centering
\includegraphics[height=8cm,width=10cm]{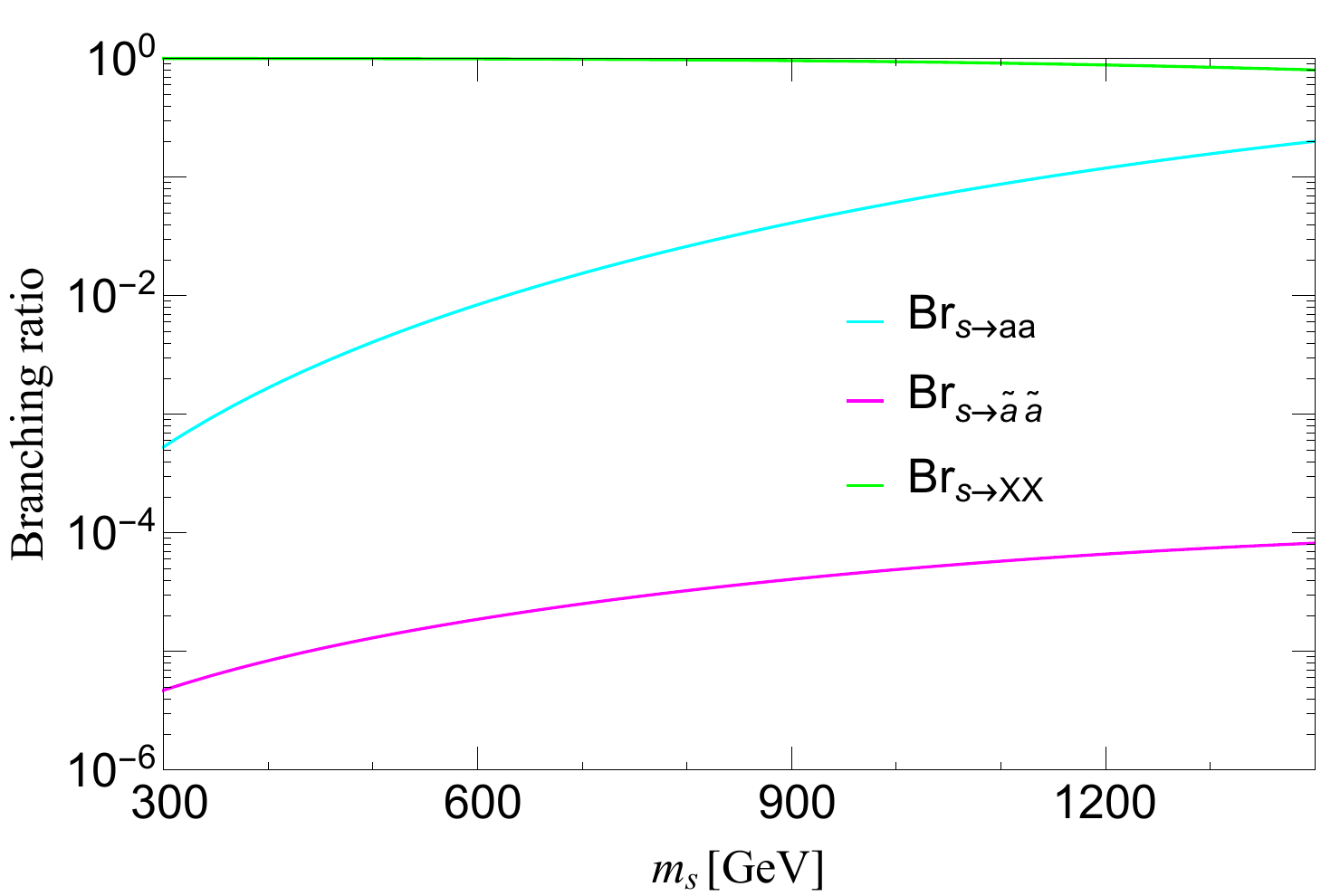}
\caption{The branching ratios of the saxion as the function of saxion mass. Here we consider the decay processes: $s \to aa$ (cyan line), $s \to \tilde{a}\tilde{a}$ (magenta line) and $s \to XX$ (green line). Other parameters are taken as
$\kappa=1$, $c_{\mu}=2$, $m_{\tilde{a}}=10$ GeV, $\mu=700$ GeV and $\mathcal{D}=4$ (The additional non-SM Higgs bosons are decoupled.)}
\label{fig2}
\end{figure}

From Eq.~\ref{kahler}, it can be seen that the self-interaction of the axion supermultiplet is induced by the K\"{a}hler potential term. Besides, the axion supermultiplet can interact with the SM particles through the Higgs field because the two Higgs doublets $H_{u}$ and $H_{d}$ carry the PQ charge. Therefore, the saxion can decay into axion pairs, axino pairs, and visible particles, whose decay widths are given by~\cite{Jeong:2012np, Bae:2013hma,Co:2016xti} 
\begin{eqnarray}
\Gamma_{s \rightarrow aa}&=&\frac{\kappa^{2} m_{s}^{3}}{64 \pi V_{PQ}^{2}},\label{saa}\\
\Gamma_{s \rightarrow \tilde{a}\tilde{a}}&=&\frac{\kappa^{2} m_{\tilde{a}}^{2} m_{s}}{8 \pi V_{PQ}^{2}},\label{sax}\\
\Gamma_{s \rightarrow XX}&=&\mathcal{D}\frac{c_{\mu}^{2} \mu^{4}}{16 \pi m_{s} V_{PQ}^{2}}. \label{sXX}
\end{eqnarray}
where $\mathcal{D}$ is the number of final state particles, such as $\mathcal{D}=4$ in the SM ($X=$ $h$, $Z$ and $W^\pm$) and $\mathcal{D}=8$ in the MSSM ($X=$ $h$, $H$, $A$, $H^\pm$, $Z$ and $W^\pm$). $c_{\mu}$ represents the PQ charge of $\mu$ term. Furthermore, we neglect the masses of final states for saxion decay to axions and axinos corresponding to Eq.~\ref{saa} and Eq.~\ref{sax}. While for Eq.~\ref{sXX} saxion decay to Higgs bosons, we take the decoupling limit regime $m_{A} \gg m_{Z}$ and the large ${\rm tan}\beta$ limit. Therefore, for $m_{s}<2m_{A}$ the kinetically allowed number of the final state Higgs is $\mathcal{D}=4$, while $\mathcal{D}=8$ for $m_{s}\gtrsim 2m_{A}$. It should be noted that the decays of heavy saxion into the gauge bosons are dominated by the decays into Goldstone modes. Thus, we can obtain the similar approximate decay widths as the Higgs final states so that the counting factor in Eq.~13 is independent of the spins of the final states. The full expressions of all decay widths can be found in the appendix of Ref.~\cite{Bae:2013hma} and are dependent on the spin of the final states. In Fig.~\ref{fig2}, we present the branching ratios of the saxion decay with saxion mass $m_{s}$ from 300 GeV to 1.4 TeV. Since we take the Higgsino mass $\mu=700$ GeV, saxion decay to Higgsino is kinetically forbidden. It can be seen that the branching ratio of $s \to XX$ is always dominant, while that of $s \to aa$ is sub-dominant in our interesting parameter space. This is because that the decay width of $s \to XX$ is greatly enhanced by the $\mu^4$. Otherwise, the branching ratio of $s \to aa$ is always larger than that of $s \to \tilde{a} \tilde{a}$ since the decay width of $s \to aa$ is proportional to $m_{s}^3$ while that of $s \to \tilde{a} \tilde{a}$ is proportional to $m_{s}$.   

\section{Saxion Cosmology and Hubble Tension}   
\label{sec3}

\begin{figure}[h]
\centering
\includegraphics[height=7cm,width=14cm]{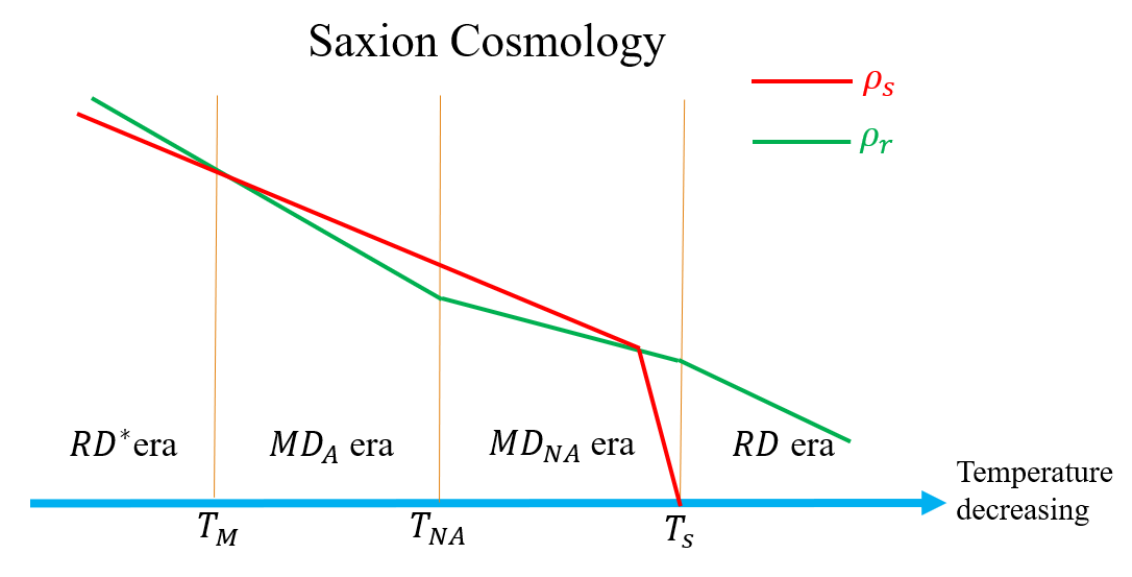}
\caption{The schematic diagram of the saxion cosmology. The red and green lines represent the saxion energy density $\rho_{s}$ and the radiation energy density $\rho_{r}$, respectively. The four eras are defined in the context.}
\label{saxion cosmology}
\end{figure}

After inflation, the saxion obtains the enormous potential $\sim m_{s}^2 s_{\rm{I}}^2$. Due to the Hubble friction, the saxion field remains fixed at $s_{I}$. However, when the Hubble parameter decreases to $3H \sim m_{s}$, the saxion field starts to oscillate at the temperature $T_{\rm osc}$, which is given by
\begin{equation}
    T_{\rm{osc}}=\left(\frac{10}{\pi^2 g_{\ast}(T_{\rm{osc}})}\right)^{1/4}\sqrt{m_{s} M_{pl}}.
\end{equation}
Here $g_{\ast}(T_{\rm{osc}})$ represents the effective number of degree of freedom in the thermal bath at the oscillation temperature $T_{\rm{osc}}$. Note that the saxion oscillation occurs at the early radiation dominated era ($RD^{\ast}$). After the oscillation of the saxion, its energy density $\rho_{s}$ drops as $a^{-3}$ as the non-relativistic matter. Since the radiation decreases as $a^{-4}$, the saxion energy density $\rho_{s}$ will eventually dominate the early universe after the temperature $T_{M}$. Assuming $g_{\ast}(T_{\rm{osc}})=g_{\ast}(T_{M})$, we derive the transition temperature $T_{M}$, which is given by
\begin{equation}
    T_{M}=3\left(\frac{10}{g_{\ast}(T_{M}) \pi^2}\right)^{1/4} \frac{m_{s}^{1/2} s_{I}^2}{M_{pl}^{3/2}}.
\end{equation}
Subsequently, the Universe enters the early matter-dominated era (MD) induced by the saxion energy density $\rho_{s}$, which consists of the adiabatic (MD$_{A}$) and non-adiabatic (MD$_{NA}$) phases. During the adiabatic era, the radiation energy density MD$_{A}$ is dominated by the initial red-shift radiation while it will be dominated by the relativistic particles from the saxion decay at the temperature $T_{NA}$. Eventually, when $H \sim \Gamma_{s}$, most of saxions will decay  at the temperature $T_{s}$, which is given by
\begin{equation}
    T_{s}=\left(\frac{90}{\pi^2 g_{\ast}(T_{s})}\right)^{1/4} \sqrt{\Gamma_{s} M_{pl}}. 
\end{equation}
After saxion decay, the Universe will enter the regular radiation-dominated era (RD). The cosmological history of the saxion is displayed in Fig.~\ref{saxion cosmology}.

Due to the saxion decay, it should be mentioned that there will be extra radiation energy from the axion, which can be injected into the early universe in the form of dark radiation. During the radiation dominant epoch, the dark radiation can alter the expansion rate of the universe via increasing the radiation energy in the Friedmann equation in a flat universe, which is related to the Hubble constant by
\begin{equation}
    H^{2}(t) \simeq \frac{8 \pi G}{3}(\rho_{\gamma}+\rho_{\nu}+\rho_{a})
\end{equation}
where $\rho_{\gamma}$, $\rho_{\nu}$ and $\rho_a$ is the energy density of photon, neutrino and axion, respectively.
The extra axion radiation will contribute to the effective number of neutrino species $ N_{\rm eff}$,
which has a positive relation to the Hubble constant. Any additional free-streaming radiation that is not included in the photon bath is usually encapsulated in the effective number of relativistic degrees of freedom $N_{\rm eff}$. In our work, we impose such constraints on the axion dark radiation by using the bounds on $\Delta N_{\rm{eff}}$ derived from the BBN~\cite{Blinov:2019gcj} and CMB~\cite{Aghanim:2018eyx}. The additional $\Delta N_{\rm eff}$ is given by 
\begin{equation}
\Delta N_{\rm eff}=N_{\rm eff}\frac{\rho_{a}}{\rho_{\nu}},
\end{equation}
where $N_{\rm eff}=3.046$ in the SM. When the neutrinos were in the thermal bath, the neutrino energy density $\rho_{\nu}$ can be related to the Standard Model energy density $\rho_{\rm SM}$ 
\begin{equation}
    \rho_{\nu}(T)=\frac{g_{\nu}(T)}{g_{\rm SM}(T)} \rho_{\rm SM}(T)
\end{equation}
where $g_{\nu}(T)$ and $g_{\rm SM}(T)$ are the number of relativistic degrees of freedom of neutrinos and Standard Model particles at the temperature $T$, respectively. Since the decay width $\Gamma_{s \rightarrow \tilde{a}\tilde{a}}$ is much smaller than $\Gamma_{s \rightarrow aa}$ and $\Gamma_{s \rightarrow {\rm XX}}$, the energy density of saxion are mostly injected in the form of axion and visible particles via saxion decay. Assuming that the additional energy density comes from the axion and the SM particles in the saxion decay, the ratio of the axion energy density to the SM energy density equals the ratio of their decay widths~\cite{Co:2016xti},
\begin{equation}
    \frac{\rho_{a} (T)}{\rho_{SM}(T)}=\frac{\Gamma_{s \to aa}}{\Gamma_{s \to XX}}
\end{equation}
Therefore, the $\Delta N_{\rm eff}$ arising from the saxion decay is given by 
\begin{equation}
\Delta N_{\rm eff}=\frac{4}{7}\frac{\Gamma_{s \rightarrow aa}}{\Gamma_{s \rightarrow {\rm XX}}} g_{\ast}\left(T_{\nu}^{\rm dec}\right),
\label{neff}
\end{equation}
where the $g_{\ast}\left(T_{\nu}^{\rm dec}\right)$ is the number of relativistic degrees of freedom at the neutrino decoupling temperature. Since both decay widths are proportional to the PQ symmetry breaking scale $1/V_{PQ}^2$, the value of $\Delta N_{\rm eff}$ is independent of $V_{PQ}$, but relies on $m_{s}$. In addition, the relation between the current Hubble constant $H_{0}$ and $H_{\rm CMB}$ inferred by CMB data is~\cite{Vagnozzi:2019ezj,Anchordoqui:2020djl}
\begin{equation}
H_{0}=H_{\rm CMB}+6.2 \Delta N_{\rm eff}
\end{equation}
with $H_{\rm CMB}=67.9$km/s/Mpc.

\begin{figure}[ht]
\centering
\includegraphics[height=8cm,width=10cm]{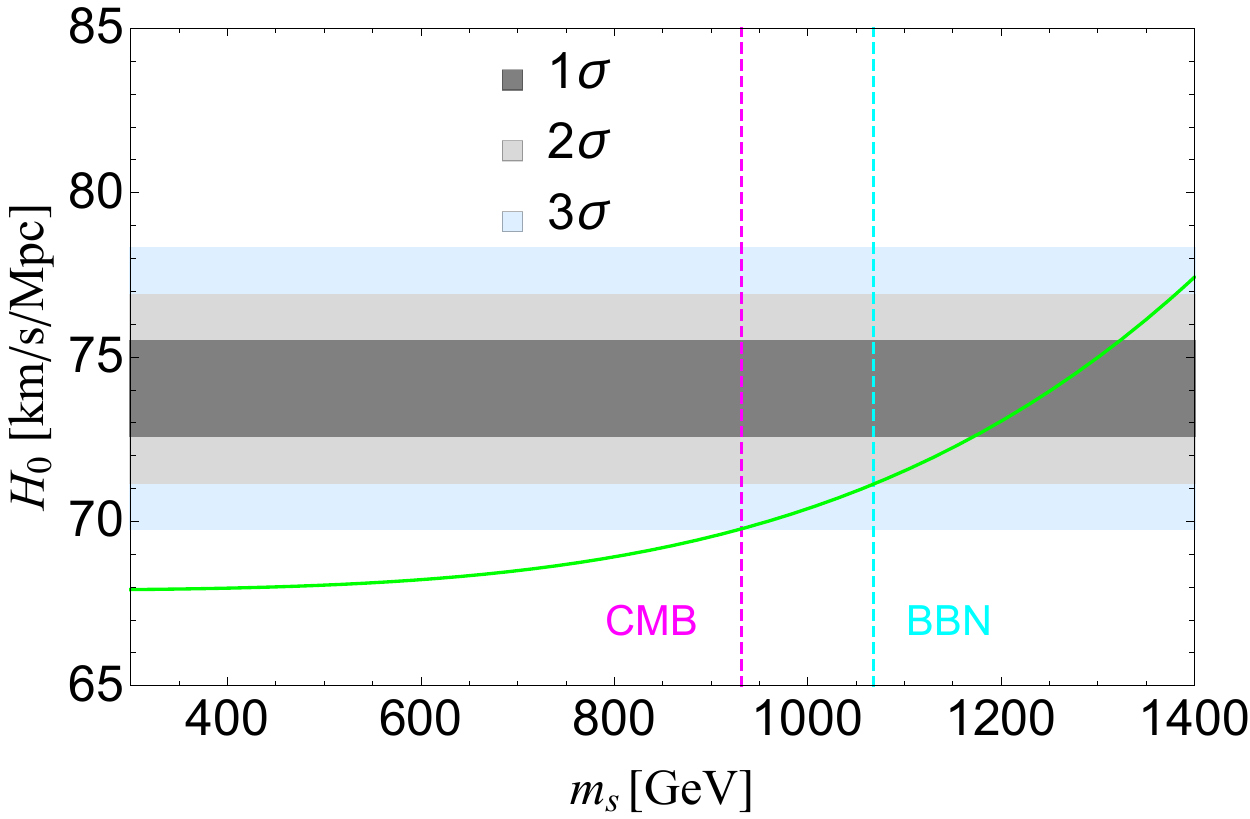}
\caption{The current Hubble constant $H_{0}$ as the function of the saxion mass $m_{s}$. The gray, light-gray and blue bands are 1$\sigma$, 2$\sigma$ and 3$\sigma$ ranges of $H_0$ for solving the Hubble tension, respectively. The magenta and cyan dashed lines are the bounds on the $\Delta N_{\rm eff}$ from CMB and BBN.}
\label{fig3}
\end{figure}

In Fig.~\ref{fig3}, we show the current Hubble constant $H_{0}$ as the function of the saxion mass $m_{s}$. Since the heavier saxion can produce more axion dark radiation, the value of $H_0$ increases with the saxion mass $m_s$. We can see that the Hubble tension will be settled within $1\sigma$ range when 1.2 TeV $< m_{s} <$ 1.3 TeV. However, it should be noted that the extra dark radiation can accelerate the expansion rate of the universe, which will result in earlier freeze-out with a larger neutron fraction and earlier onset of nucleosynthesis. Furthermore, this eventually leads to a larger $^4$He abundance, destroying the perfect predictions of standard BBN theory. In addition, the extra radiation also delays the matter-radiation equality, altering the observed CMB spectrum. Therefore, the extra effective number of neutrino species $\Delta N_{\rm eff}$ arising from the extra dark radiation is strongly constrained by both the CMB and the BBN, which further produce a constraint on the saxion mass via Eq.~\ref{neff}. It can be seen that the current BBN bound $\Delta N_{\rm eff}<0.5$~\cite{Blinov:2019gcj} is weaker than the CMB and large scale structure bound $\Delta N_{\rm eff}<0.28$~\cite{Aghanim:2018eyx}. But the BBN limit is not as sensitive to the choice of the cosmological model. Under this constraint, the Hubble tension is only marginally reduced to $2\sigma$ via simply increasing $\Delta N_{\rm eff}$. While the currently allowed range $10^{-2}<\Delta N_{\rm eff}<0.5$ will be within the reach of CMB-S4 predictions~\cite{Abazajian:2016yjj}. Besides, the axions from saxion decay are relativistic so that there must exist a strong constrain from the free-streaming length. The comving free-streaming length of the axion generated by saxion decay can be calculated as~\cite{Lee:2014xua}
\begin{eqnarray}
    \lambda_{\rm fs} & \equiv & \int_{\tau_{\rm s}}^{t_{\rm eq}} \frac{v_{a} dt}{a(t)}=\int_{\tau_{\rm s}}^{t_{\rm nr}} \frac{dt}{a(t)} + \int_{t_{\rm nr}}^{t_{\rm eq}} \frac{v_{a} dt}{a(t)} \\ \nonumber
    & \approx & \frac{1}{H_{0}} \left(\frac{H_{0}}{\Gamma_{s}}\right)^{1/2}\left(\frac{m_{s}/2}{m_{a}}\right) \left(\frac{T_{\rm eq}}{T_{0}}\right)^{1/4} \left\{ 1+\frac{1}{2} {\rm ln}\Bigg[ \frac{\Gamma_{s}}{H_{0}} \left(\frac{m_{a}}{m_{s}/2}\right)^2 \left(\frac{T_{0}}{T_{\rm eq}}\right)^{3/2} \Bigg]\right\}
\end{eqnarray}
where $\tau_{s}$ is the lifetime of saxion, $a(t)$ is the scale factor and $v_{a}$ is the velocity of axion. $t_{\rm eq}$ is the time of matter-radiation equality with the temperature $T_{\rm eq}$. $t_{\rm nr}$ is the time when the axion from saxion decay is non-relativistic. $\Gamma_{s}$ is the total decay width of saxion and $m_{a}$ is the axion mass. The ratio of the temperature $T_{\rm eq}$ to $T_{0}$ is about 3200. As mentioned before, the $\Delta N_{\rm eff}$ is independent of $V_{PQ}$, but only relies on the saxion mass $m_{s}$. When calculating the free-streaming length $\lambda_{\rm fs}$, we set the saxion mass $m_{s}=10^3$ GeV, which alleviates the Hubble tension to 3$\sigma$. Also, other relevant parameters are the same as in Fig.~\ref{fig2}. Therefore, we can derive the bounds of free-streaming length on $V_{PQ}$ and axion mass $m_{a}$, which is given by, 
\begin{equation}
    \lambda_{\rm fs} \simeq 0.6 \,{\rm Mpc} \left(\frac{V_{PQ}}{10^{13} \, {\rm GeV}}\right) \left(\frac{35 \, {\rm keV}}{m_{a}}\right) \left\{1 + \frac{1}{2} {\rm ln}\Bigg[14494 \left(\frac{10^{13} \, {\rm GeV}}{V_{PQ}}\right)^2 \left(\frac{m_{a}}{35 \, {\rm keV}}\right)^2 \Bigg]\right\}
\end{equation}
In our following calculation, we take into account the free-streaming constraint $\lambda_{\rm fs} \lesssim 0.6$ Mpc.

\section{Late Decaying DM: Gravitino/Axino}
\label{sec4}

In supersymmetric axion models, the gravitino/axino can play the role of dark matter particle. The gravitino mass $m_{3/2}$ depends on the scheme of supersymmetry breaking~\cite{Pagels:1981ke,Weinberg:1982zq,Khlopov:1984pf,Dine:1994vc,Giudice:1998bp,Randall:1998uk,Ellis:2003dn,Buchmuller:2005rt,Dudas:2017rpa}, which can be at the same order of the axino mass $m_{\tilde{a}}$~\cite{Goto:1991gq,Chun:1992zk,Chun:1995hc,Kim:2012bb}. Therefore, they can consist of a two-component dark matter sector, where the heavier one is the late decaying DM and the lighter one is the LSP dark matter. However, if the mass of the gravitino is heavier than keV, the abundance of the gravitino will easily overclose the universe. To give the present observed dark matter relic density, the reheating temperature $T_{R}$ after inflation is usually much lower than that for the thermal leptogenesis. Fortunately, in our model, the saxion decay can help to relax the above cosmological constraint on the gravitino mass and the reheating temperature $T_{R}$.


Due to saxion decay, a large amount of the entropy was injected into the universe. Thus the dark matter ($\tilde{G}$ and $\tilde{a}$) relic density produced before can be diluted~\cite{Banks:2002sd,Kawasaki:2008jc}. The effect of dilution can be parameterized by the factor $D_{s}$, which is given by
\begin{equation}
    D_{s}=\left(\frac{9 g_{\ast}(T_{s})}{g_{\ast}(T_{M})}\right)^{1/4} \frac{m_{s}^{1/2} s_{I}^{2}}{\sqrt{\Gamma_{s}} 
        M_{pl}^{2}},
\end{equation}
where $g_{\ast}(T_{s})=10.75$ and $g_{\ast}(T_{M})=228.75$ are the number of relativistic degrees of freedom at the temperature $T_{s}$ and $T_{M}$ respectively. 

The relic density of the gravitino depends on whether it is in the thermal bath~\cite{Fujii:2002fv}. The freeze-out temperature of the gravitino is given by
\begin{equation}
    T^{f}_{3/2} \approx 10^{13} {\rm GeV} \left(\frac{g_{\ast}}{230}\right)^{\frac{1}{2}} \left(\frac{m_{3/2}}{1 {\rm GeV}}\right)^{2} \left(\frac{1 {\rm TeV}}{m_{\tilde{g}}}\right)^{2},
\end{equation}
where $m_{\tilde{g}}$ is the mass of gluino. $g_{\ast}$ is the effective degrees of freedom of relativistic particles when the gravitinos are out of equilibrium. If the reheating temperature $T_{R}>T^{f}_{3/2}$, the pre-existing gravitinos are in the thermal equilibrium due to the high reheating temperature $T_{R}$ and then freeze out as the expansion of the universe. The undiluted freeze-out gravitino yield at the equilibrium
\begin{equation}
    Y^{FO}_{3/2}=\frac{135 \zeta\left(3\right) g_{3/2}}{8 \pi^{4}g_{\ast}},
\end{equation}
where $g_{3/2}=4$ is the internal degrees of freedom of the gravitino, $\zeta$ is the zeta function. On the contrary, when $T_{R}<T_{3/2}^f$, the pre-existing gravitinos may be out of the thermal equilibrium. In this case, the gravitino can also be produced by the scattering processes, such as $\tilde{g}\tilde{g} \to \tilde{G}\tilde{G}$, which is the so-called freeze-in mechanism. The resulting undiluted freeze-in gravitino yield can be obtained by
\begin{equation}
    Y^{FI}_{3/2}=\sum^{3}_{i=1} y_{i} g^{2}_{i}\left(T_{R}\right) \left(1+\frac{M^{2}_{i}\left(T_{R}\right)}{3m^{2}_{3/2}}\right) \times \ln\left(\frac{k_{i}}{g_{i}\left(T_{R}\right)}\right) \left(\frac{T_{R}}{10^{10} {\rm GeV}}\right)
\end{equation}
where $g_{i}$ and $M_{i}$ are the gauge coupling constants and the gaugino masses for $M_{1}=M_{2}=500$GeV, $M_{3}=1$TeV. The constants $c_{i}$, $k_{i}$ and $y_{i}$ are associated with the Standard Model gauge group $SU(3)_{c} \times SU(2)_{L} \times U(1)_{Y}$~\cite{Pradler:2006hh}. Thus, the total undiluted gravitino abundance depends on the relation between reheating temperature $T_{R}$ and freeze-out temperature $T_{3/2}^f$, which can be written as
\begin{equation}
\Omega_{3/2}h^2=m_{3/2}  s\left( T_{0}\right) h^{2}/ \rho_{c} \left[Y^{FO}_{3/2}\Theta(T_{R}-T_{3/2}^{f})+Y^{FI}_{3/2} \Theta(T_{3/2}^{f}-T_{R}))\right].
\end{equation}
where $h$ is the Hubble constant in the units of 100km/s/Mpc. $s\left(T_{0}\right)$ is the current entropy density and $\rho_{c}$ is the critical density with $\rho_{c}/\left[s\left(T_{0}\right)h^2\right]=3.6 \times 10^{-9} {\rm GeV}$. It should be mentioned that the Eq.~27 is only applicable for $T_{R} \ll T_{f}$ or $T_{R} \gg T_{f}$, which corresponds to the freeze-in or freeze-out production, respectively. If the reheating temperature $T_{R}$ is comparable with $T_{f}$, one need to numerically solve the Boltzmann equation to obtain the dark matter relic density.

\begin{figure}[ht]
\centering
\includegraphics[height=6.5cm,width=6.5cm]{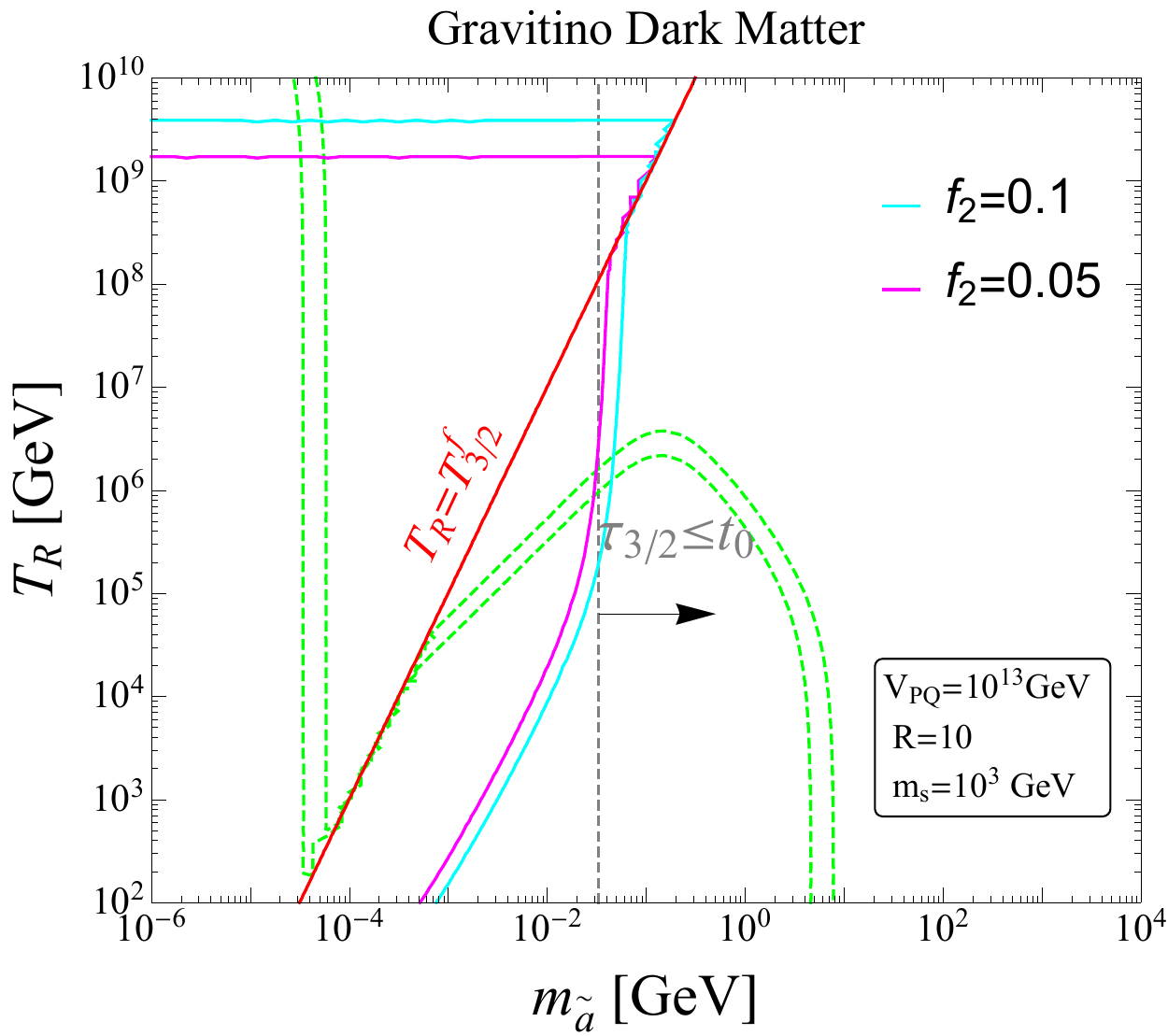}
\hspace{0.5cm}
\includegraphics[height=6.5cm,width=6.5cm]{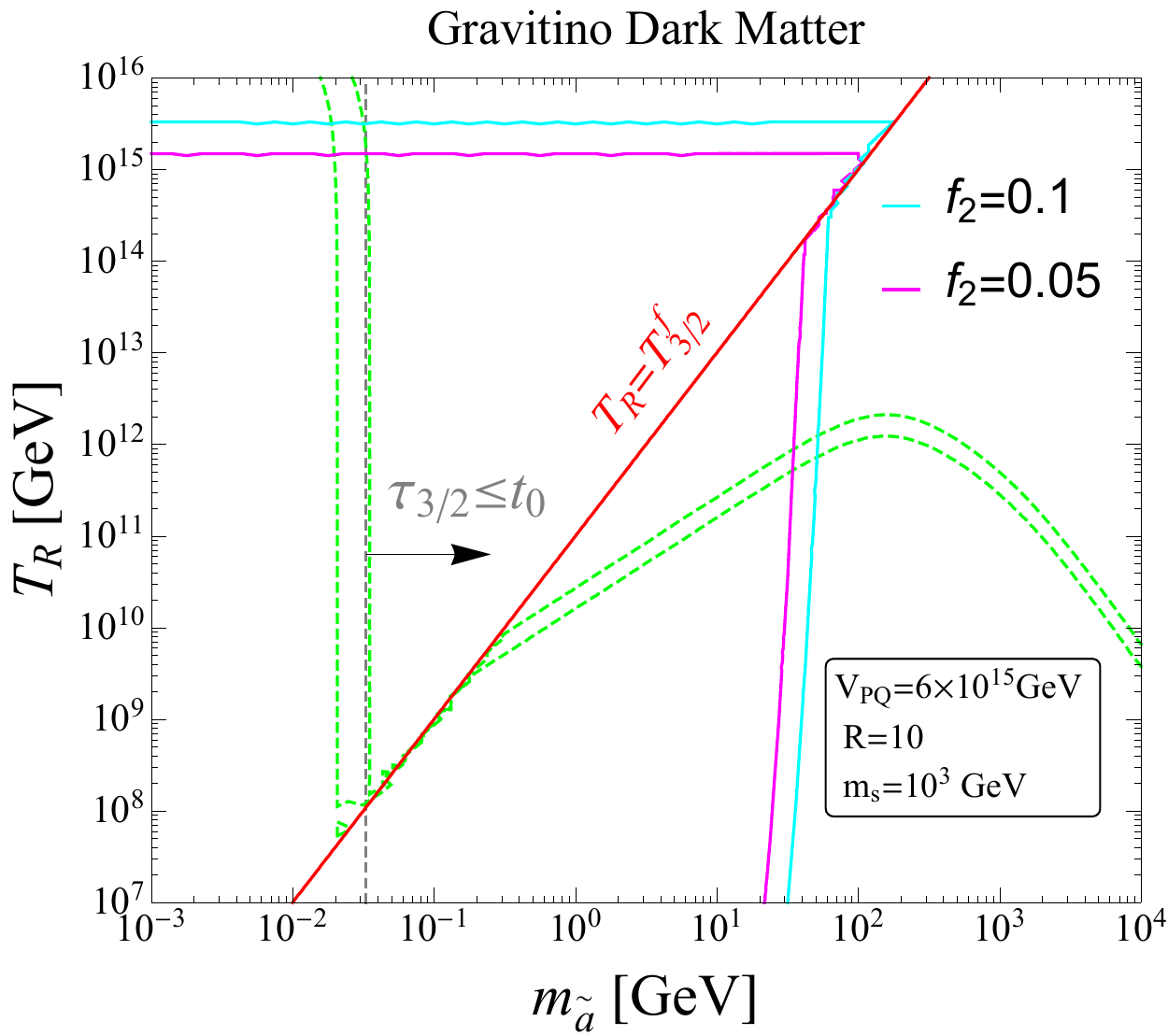}
\includegraphics[height=6.5cm,width=6.5cm]{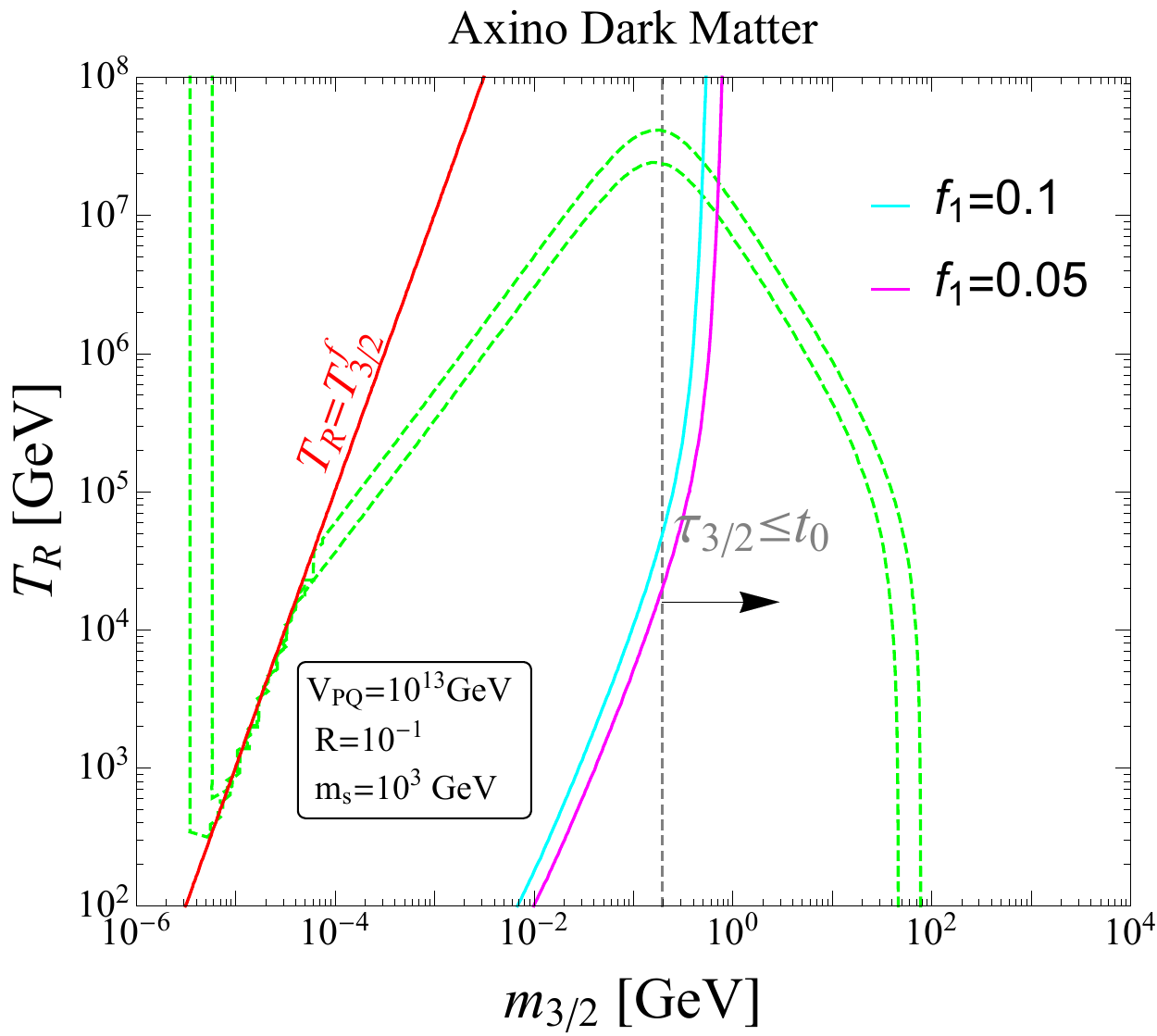}
\hspace{0.5cm}
\includegraphics[height=6.5cm,width=6.5cm]{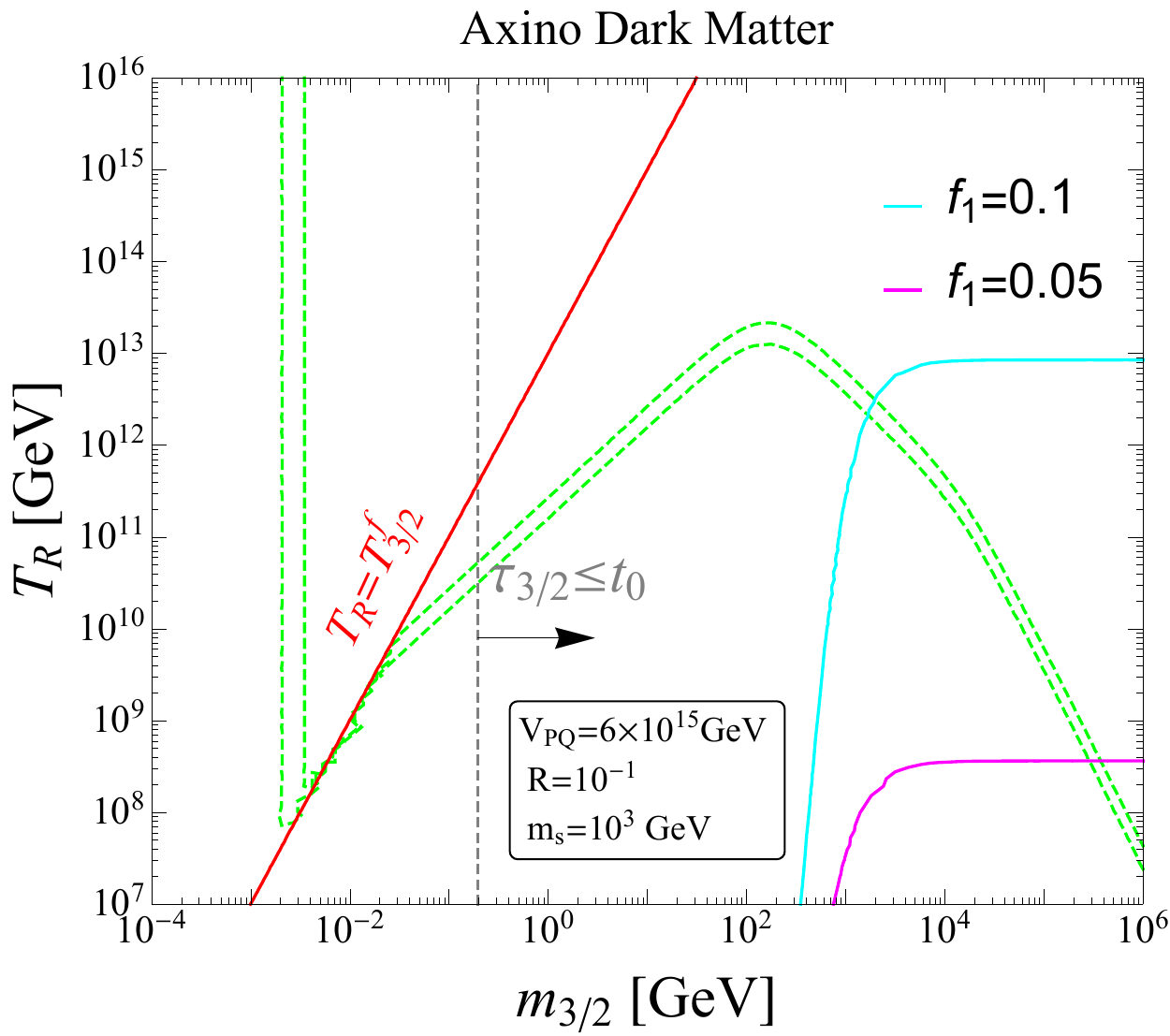}
\caption{The constraint of DM relic density $0.075<\Omega_{\rm DM} h^2<0.126$ (green band) for the gravitino LSP (upper panel) and the axino LSP (lower panel) on the plane of the reheating temperature $T_R$ versus their masses $m_{3/2}$ and $m_{\tilde{a}}$. $f_{1}=\Omega_{3/2}h^2/\Omega_{\rm DM} h^2$ and $f_{2}=\Omega_{\tilde{a}}h^2/\Omega_{\rm DM} h^2$ are the fraction of the energy density of the gravitino and axino NLSP in the total DM energy density, respectively. We assume the saxion mass $m_{s}=10^3$GeV. The right side of the gray dash line corresponds to the lifetime of the NLSP $\tau_{\rm NLSP} < t_{0}$, where $t_{0}=13.7$Gyr is the lifetime of the universe. While the region above the red line denotes the reheating temperature $T_R$ is higher than the freeze-out temperature of the gravitino $T_{3/2}^f$.}
\label{relic density}
\end{figure}

Similarly, if the axino is the LSP, the corresponding freeze-out temperature of the axino $T_{\tilde{a}}^f$ is calculated by, 
\begin{equation}
    T^{f}_{\tilde{a}}=10^{11} {\rm GeV} \left(\frac{\sqrt{2} V_{PQ}}{N_{\rm DW}10^{12}{\rm GeV}}\right)^{2} \left(\frac{0.1}{\alpha_{s}}\right)^{3},
\end{equation}
where $N_{\rm DW}=6$ for SUSY DFSZ model is the domain wall number and $\alpha_{s}$ is a strong interaction coupling constant. Provided that the reheating temperature $T_{R}$ is higher than the freeze-out temperature $T_{\tilde{a}}^f$, the effect of $T_{R}$ will compensate the feeble interaction between the axino and other particles, and axino is in thermal equilibrium. Then the undiluted yield of the axino in the thermal bath is given by
\begin{equation}
    Y^{FO}_{\tilde{a}}=\frac{135 \zeta\left(3\right) g_{\tilde{a}}}{8 \pi^{4}g_{\ast}},
\end{equation}
where the internal degrees of freedom $g_{\tilde{a}}=2$ is for the axino. On the other hand, when the axino is out of  equilibrium, its undiluted yield arising from the scattering of gluon and gluino is estimated by~\cite{Strumia:2010aa}
\begin{equation}
    Y^{FI-{\rm scattering}}_{\tilde{a}} \simeq 2 \times 10^{-6} {\rm ln}\left(\frac{3}{g_{3}}\right)\left(\frac{N_{DW}}{6}\right)^2\left(\frac{T_{R}}{10^{10}{\rm GeV}}\right)\left(\frac{10^{14}{\rm GeV}}{V_{PQ}}\right)^2,
\end{equation}
where $g_{3}$ is the strong gauge coupling. In addition to the scattering of gluino and gluon, the axino can also be produced by Higgsino decay~\cite{Chun:2011zd,Bae:2011jb,Co:2015pka} and the yield is expressed by
\begin{equation}
    Y_{\tilde{a}}^{FI-{\rm decay}} \approx 2.9\times 10^{-4} q_{\mu}^2\left(\frac{230}{g_{\ast}}\right)^{3/2}\left(\frac{\mu}{300 {\rm GeV}}\right)\left(\frac{10^{10} {\rm GeV}}{V_{PQ}}\right)^2
\end{equation}
where we also take $\mu=700$GeV. The axino abundance produced from Higgsino decay is comparable with that resulting from scattering at low reheating temperature $T_{R}$. While the axino yield produced from scattering is always dominated at high reheating temperature $T_{R}$, which allows us to ignore the axino yield produced from Higgsino decay. The total undiluted axino abundance generated by the Freeze-in mechanism is composed by
\begin{equation}
    Y_{\tilde{a}}^{FI}=Y_{\tilde{a}}^{FI-{\rm scattering}}+Y_{\tilde{a}}^{FI-{\rm decay}}
\end{equation}
Therefore, the undiluted relic abundance of the axino can be written as
\begin{equation}
    \Omega_{\tilde{a}} h^2=m_{\tilde{a}} s\left(T_{0}\right) h^{2}/ \rho_{c}\left[Y^{FI}_{\tilde{a}}\Theta(T_{\tilde{a}}^{f}-T_{R})+Y^{FO}_{\tilde{a}} \Theta(T_{R}-T_{\tilde{a}}^{f}) \right].
\end{equation}

Assuming that the axino and gravitino were produced before $\rho_{s}$ dominated the universe, both will be diluted by the saxion decay. Moreover, we require that the lifetime of the NLSP is longer than the age of the universe ($t_0$). As a result, the total DM relic density is
\begin{equation}
    \Omega_{DM} h^{2}= \frac{1}{D_{s}} \left(\Omega_{3/2}h^{2}+\Omega_{\tilde{a}}h^{2}\right).
\end{equation}
In Fig.~\ref{relic density}, we show the constraint of DM relic density $0.075<\Omega_{\rm DM} h^2<0.126$ for the gravitino LSP and the axino LSP, respectively. In the region above the red line, the gravitino is in the thermal equilibrium and is produced by the freeze-out mechanism. While in the region below the red line, it never reaches the thermal equilibrium and is produced by the freeze-in mechanism. It should be mentioned that the axino never reaches the thermal equilibrium in our plots because the axino freeze-out temperature $T_{a}^f$ is larger than the reheating temperature $T_{R}$. With the increase of the reheating temperature $T_{R}$, the large dilution effect is required to avoid overclosing the universe. In other words, the strong constraint on $T_{R}$ from the observed DM relic density can be relaxed by dilution. Since the high $V_{PQ}$ can produce the large dilution factor, the reheating temperature $T_R$ in the right panel is much larger than that in the left panel. In addition, for a given mass ratio $R$, when the gravitino mass $m_{3/2}$ becomes small, the freeze-out temperature of the gravitino $T_{3/2}^f$ is lower than that of the axino $T_{\tilde{a}}^f$. On the other hand, when $m_{3/2}$ becomes heavy, $T_{3/2}^f$ will be larger than $T_{\tilde{a}}^f$. Therefore, the gravitino will dominate the DM relic density in the small $m_{3/2}$ region because of the small freeze-out temperature. While the axino will be the main component of the DM abundance in the large $m_{3/2}$ region. In the upper panel, we also find that the gravitino DM yield $Y_{3/2}^{FO} $ will not change with the variation of the reheating temperature $T_{R}$ when the gravitino is in the thermal bath. This makes the fraction $f_{2}$ larger with the increase of high $T_{R}$ because more axinos are produced. While for the axino DM in the lower panel, it can be also seen that there is a tension between the lifetime of the decaying gravitino and the fraction of it in the total DM energy density that is required to be less than 10\% by the measurements of the CMB and matter power spectrum~\cite{Poulin:2016nat}. Therefore we will focus on the gravitino DM plus the axino decaying DM in the following study.


\section{Axion from Late Decaying DM in Hyper-Kamiokande}
\label{sec5}

Owing to the mass splitting between the NLSP and LSP, the NLSP can decay into the LSP plus an axion in our model. As known, the differential events caused by the absorption of the axion always depend on the energy differential flux of the axion arriving at the earth, $d\Phi_{a}/dE_{a}$, which include the contributions in the galaxy and outside the galaxy. The energy spectrum of the axion produced from the late decaying DM is given by, 
\begin{equation}
    \frac{dN}{dE_{a}}=N_{a}\delta(E_{a}-E_{\rm em})
\end{equation}
where $E_{\rm em}$ is the energy of axion at emission and $N_{a}$ is the number of axion in the final state. Assuming that the parent particle $X$ is at rest, we can calculate $E_{\rm em}$ by using four-momentum conservation in the massless limit $m_{a}=0$,
\begin{equation}
    E_{\rm em}=\frac{m_{X}}{2}(1-y^2)
\end{equation}
where $y=R$ for the gravitino NLSP and $y=R^{-1}$ for the axino NLSP. 

The differential flux of axion from DM decay within the galaxy arriving at earth is obtained by the line of sight integral,
\begin{equation}
    \frac{d\Phi_{gal}}{dE_{a}}=\frac{f Br_{a} e^{-t_{0}/\tau_{X}}}{\tau_{X} m_{X}}\frac{dN}{dE_{a}} \times R_{\rm sol} \rho_{\rm sol} \mathcal{J}
\label{galactic}
\end{equation}
where $R_{\rm sol}=8.33$ kpc is the distance to the galaxy center and $\rho_{\rm sol}=0.3$ GeV$^3/$cm is the DM energy density at the position of the earth; $t_{0}=13.7$ Gyr and $\tau_{X}$ is the lifetime of universe and late decaying DM respectively. $Br_{a}$ is the branching ratio of $X$ decay to axions. Same as defined in Fig.~\ref{relic density}, $f$ is the fraction of the decaying DM to the total DM energy density, which has to be less than about 10\% as the constraints of CMB and matter power spectrum~\cite{Poulin:2016nat}. In addition, assuming that $\mathcal{J}$ is independent of the angle and direction, we average the $\mathcal{J}$ over all directions and take the value of $\mathcal{J}=2.1$ derived from the NFW density profile~\cite{Navarro:1995iw}. Note that we take a 5\% Gaussian distribution to display the differential galactic flux.

\begin{figure}[h]
\centering
\includegraphics[height=7cm,width=7cm]{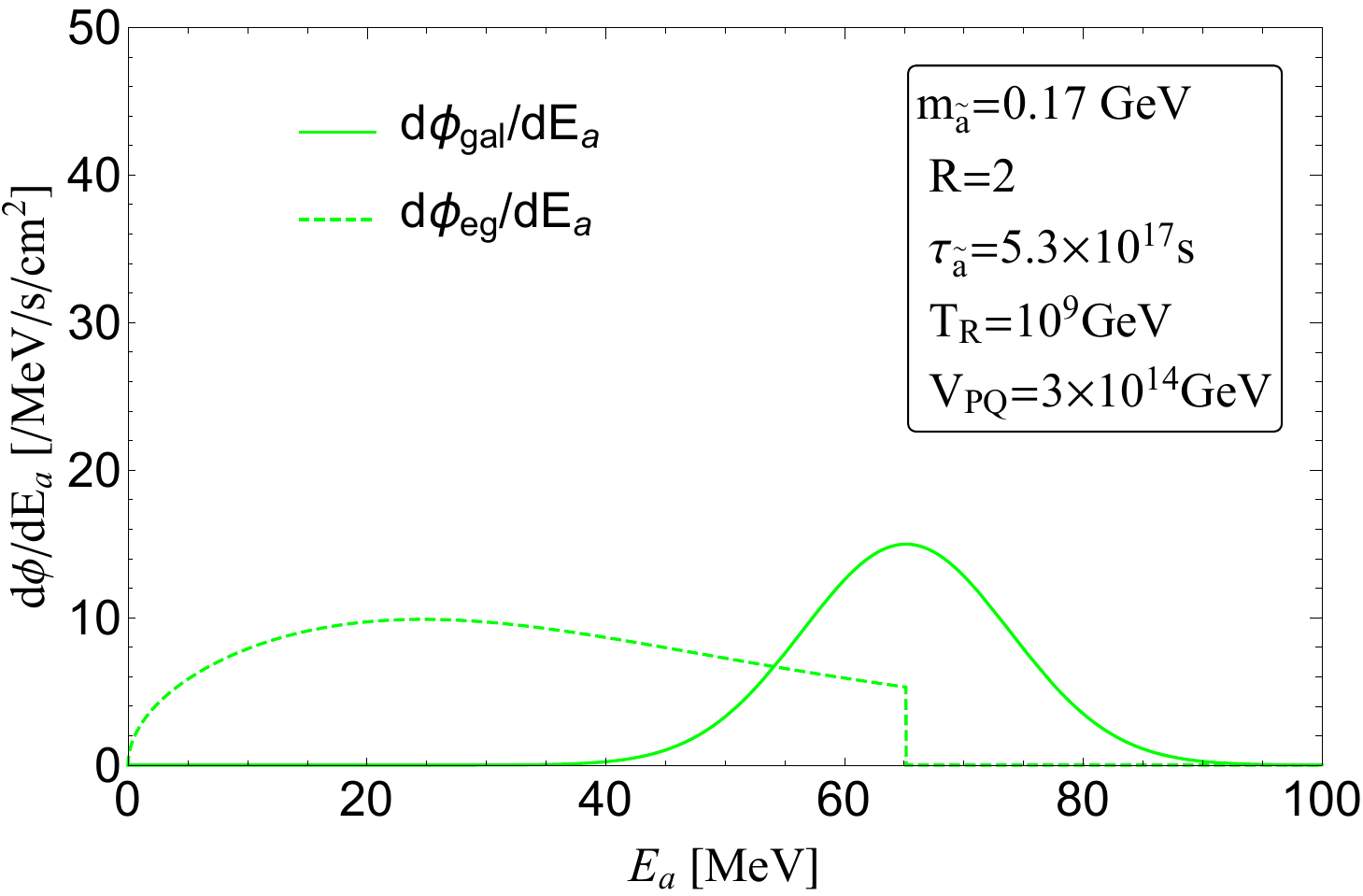}
\includegraphics[height=7cm,width=7cm]{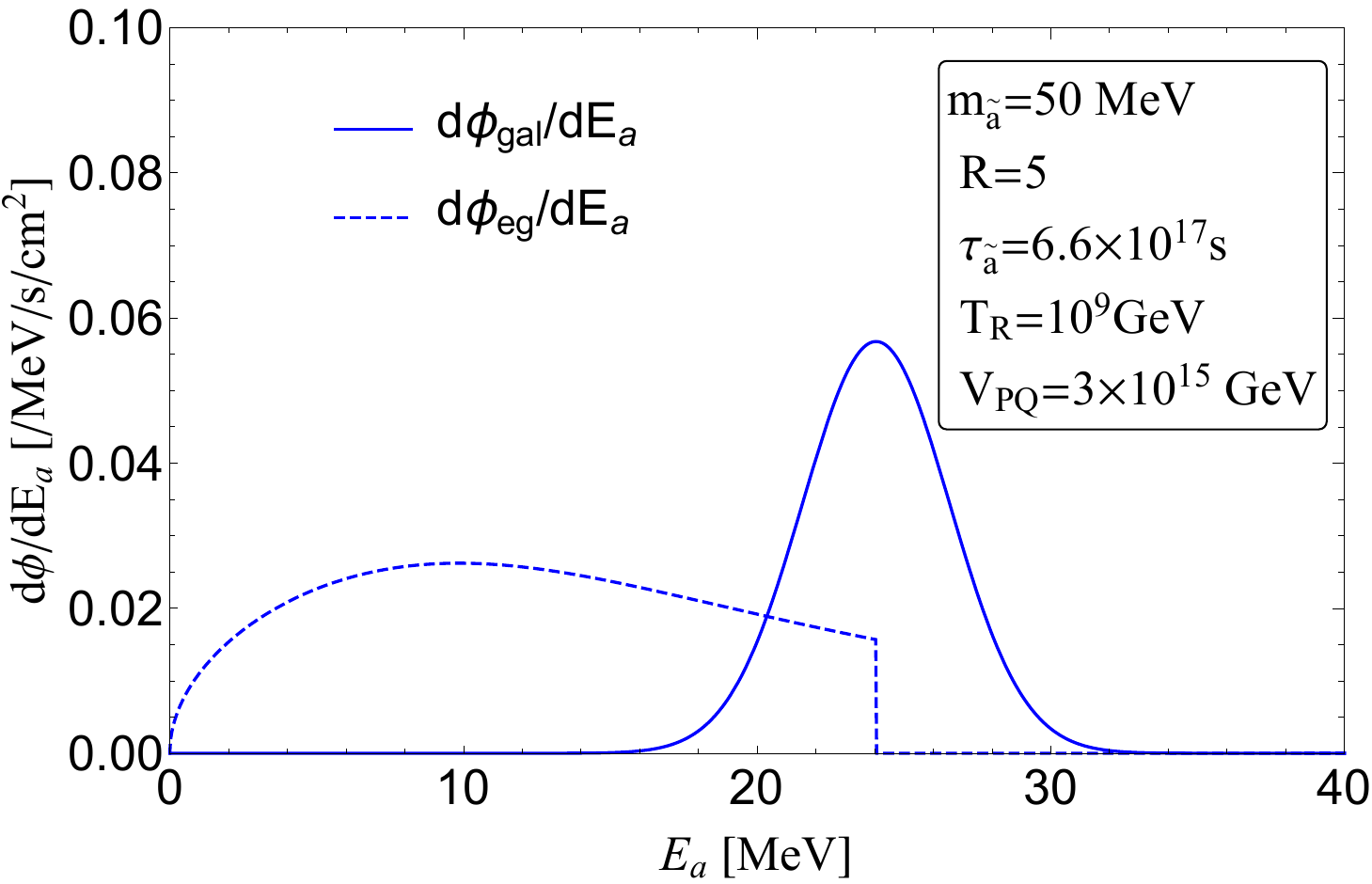}
\includegraphics[height=7cm,width=7cm]{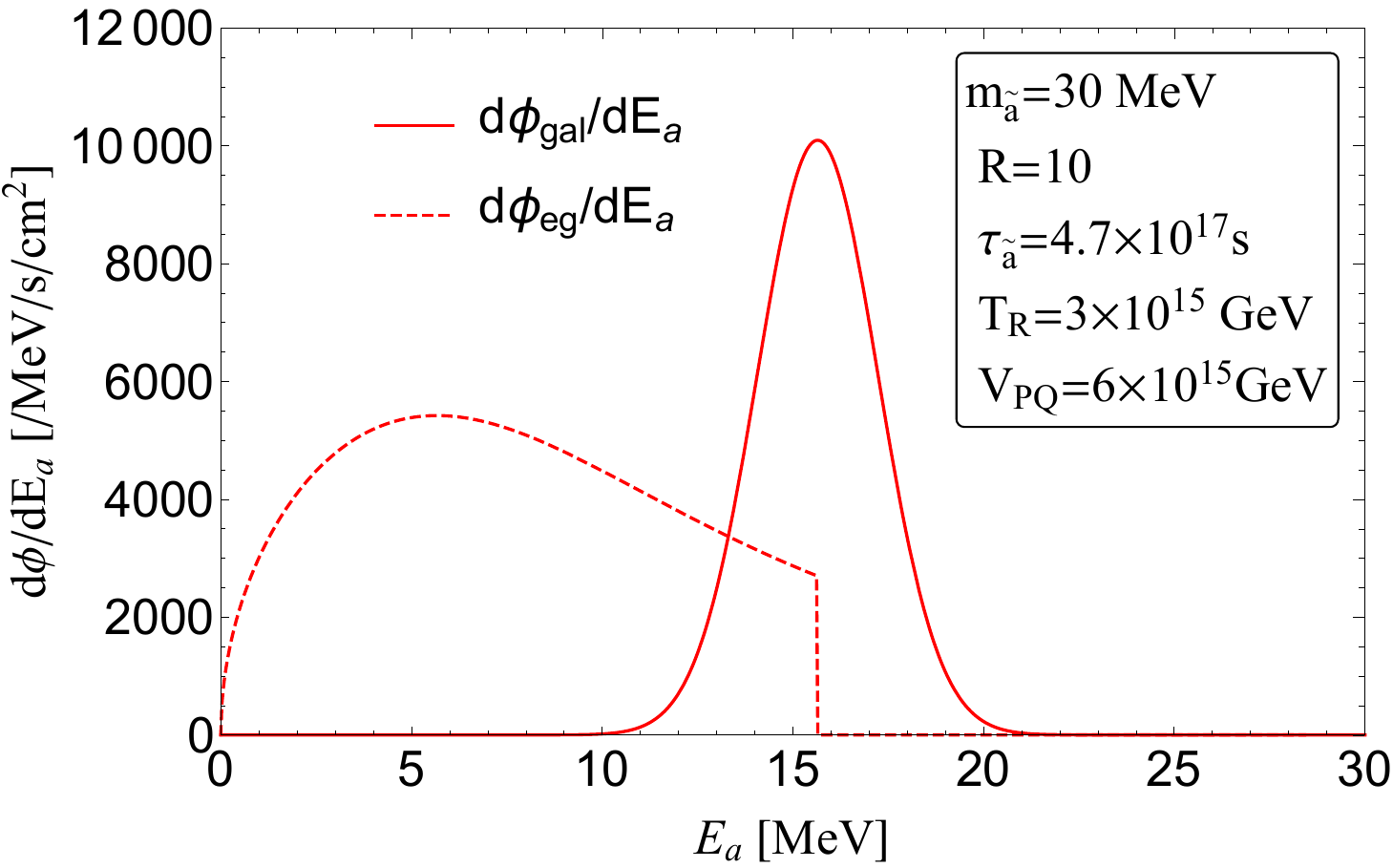}
\includegraphics[height=7cm,width=7cm]{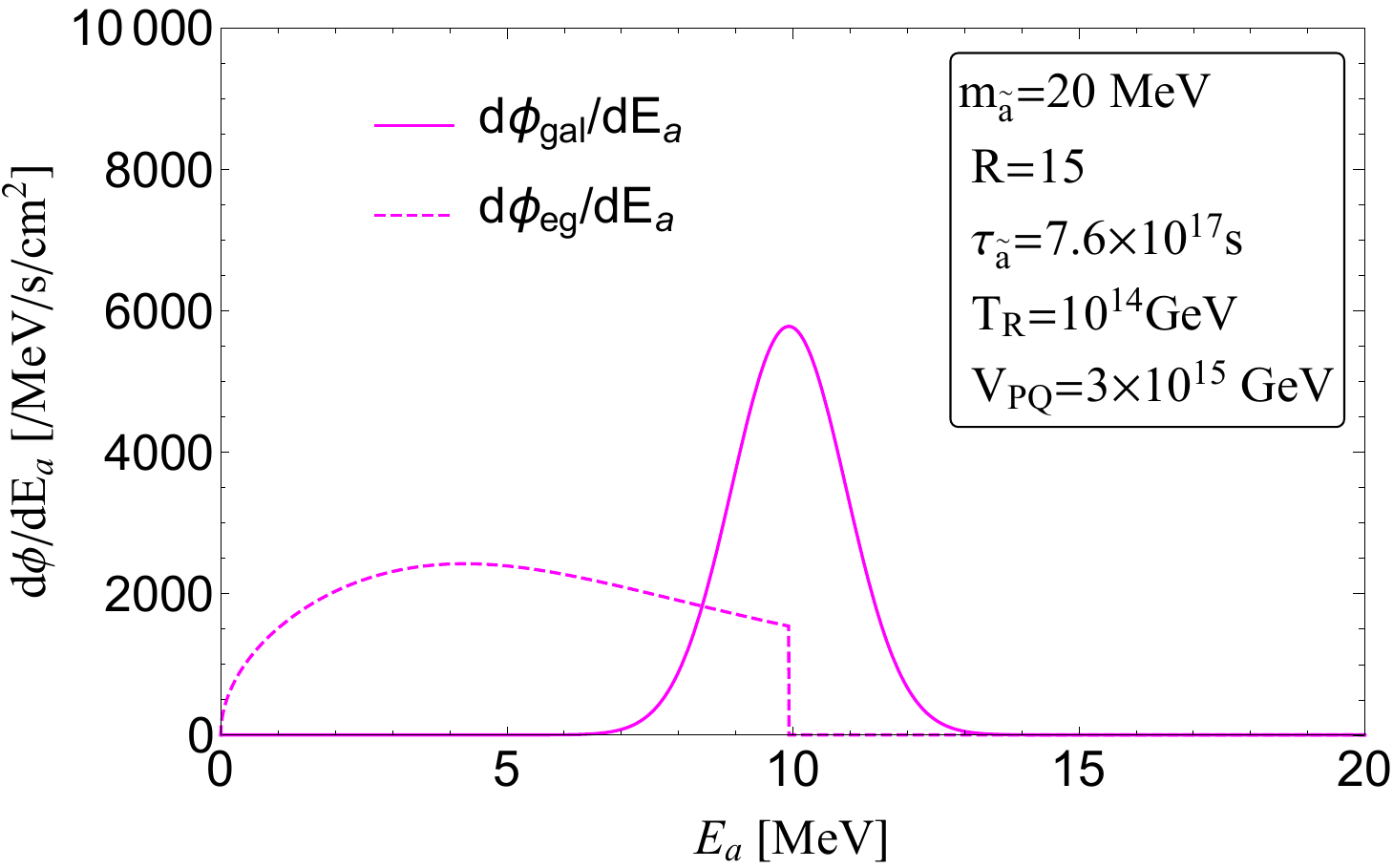}
\caption{The dependence of the flux of supersymmetric axion dark radiation on the axion kinetic energy in Milky Way galaxy and outside galaxy.}
\label{flux}
\end{figure}
In the calculation of the differential flux from extra-galactic distances, one should take the red-shift effect into account. Since the momentum of the axion is inversely proportional to the scale factor $a$, the relation between the momentum at emission $p_{\rm em}$ and that absorbed by detector $p_{a}$ is related by the red-shift factor $z$,
\begin{equation}
p_{\rm em}(z)=(1+z) \times p_{a}.
\end{equation}
Besides, the cosmic time at redshift $z$ for a flat universe is obtained

\begin{equation}
    t(z)=\frac{1}{3 H_{0} \sqrt{\Omega_{\Lambda}}}{\rm ln} \bigg[\frac{\sqrt{1+(\Omega_{m}/\Omega_{\Lambda})(1+z)^3}+1}{\sqrt{1+(\Omega_{m}/\Omega_{\Lambda})(1+z)^3}-1}\bigg],
\end{equation}
where $H_{0}$ is the current Hubble constant; $\Omega_{m}=0.315$ is the ratio of matter energy density $\rho_{m}$ to the critical density $\rho_{c}=3 M_{P}^2 H_{0}^2$, while $\Omega_{\Lambda}=1-\Omega_{m}$ is the ratio of cosmological constant $\rho_{\Lambda}$ to $\rho_{c}$. As a result, the extra-galactic flux for a 2-body decay is given by
\begin{equation}
    \frac{d\Phi_{eg}}{dE_{a}}=\frac{N_{a}}{p_{a}}\frac{f Br_{a} \Omega_{\rm dm}\rho_{c}}{\tau_{X} m_{X} H_{0}}\frac{e^{-t(z)/\tau_{X}}}{\sqrt{(1+z)^3\Omega_{m}+\Omega_{\Lambda}}} \Theta(z),
\label{extragalactic}
\end{equation}
where $\Omega_{\rm dm}=0.2607$ is the current DM density~\cite{Aghanim:2018eyx}. The total differential flux $d\Phi_{a}/dE_{a}$ consists of the galactic and the extra-galactic flux,
\begin{equation}
   \frac{d\Phi_{a}}{dE_{a}}= \left(\frac{d\Phi_{gal}}{dE_{a}}+\frac{d\Phi_{eg}}{dE_{a}}\right)
\end{equation}
From Eq.~\ref{galactic} and~\ref{extragalactic}, it is easy to see that the flux of the axion radiation depends on the lifetime $\tau_{NLSP}$ of the NLSP, the mass of NLSP $m_{X}$ and the fraction $f$. The light NLSP mass and large fraction can lead to a large flux. However, for a given mass ratio $R$, a lighter NLSP mass it is, a longer NLSP lifetime it has. Such a property will suppress the flux of axion radiation for lighter NLSP.
Besides, if the NLSP is the gravitino, the late decaying DM will dominate the DM relic density, which is excluded by the CMB constraint. Hence, we only consider the axino NLSP as the late decaying DM.

It should be noted that the kinetic energy of the axion from the axino decay depends on our model parameters. When the kinetic energy lies in the keV range, the flux of our axion will be much smaller than that of the solar axion~\cite{Derbin:2011gg,Arisaka:2012pb} so that we cannot obtain the sensitivity of our axion radiation in the solar axion experiments. Therefore, we focus on the NLSP mass $m_{NLSP}$ is larger than keV scale, such as the MeV axino, which will produce the MeV axion from the axino decay. 

In Fig.~\ref{flux}, we show the differential flux of the axion radiation. The relevant parameters, such as the life-time $\tau_{\tilde{a}}$, the mass $\tilde{a}$, the mass ratio $R$, the reheating temperature $T_R$ and the PQ symmetry breaking scale $V_{PQ}$, are chosen to satisfy the aforementioned constraints. As above discussed, the flux strongly relies on the fraction of late-decaying DM, like in the left-bottom panel, which reaches about 10\% and has a much larger flux other three cases. Besides, we find that the galactic contribution is slightly smaller than the extra-galactic contribution. Due to the red-shift effect, the extra-galactic differential flux spreads and ends at the corresponding axion kinetic energy $E_{\rm em}$ in each panel.

In order to detect such MeV axions, we consider the inverse Primakoff scattering process $a+A \to \gamma +A$, in which the axion scatters off with the atom $A$ induced by the axion-photon coupling~\footnote{On the other hand, the axion can also be absorbed through the axion-electron interaction. By calculating the number of such events in our scenario as the Ref.~\cite{Cui:2017ytb}, we found that the resulting bound on $g_{aee}$ is much weaker than that from the solar axion because the flux of the axion in our model is several orders magnitude smaller than that of the solar axion.}. Since the inelastic scattering cross section is suppressed by the atomic number $Z$, we study the elastic scattering process. The corresponding elastic cross section is given by~\cite{Abe:2021ocf} 
\begin{equation}
    \sigma_{\rm el}\left(E_{a}\right)= \frac{\alpha Z^2 g_{a \gamma \gamma}^2}{2} \bigg[{\rm ln}\left(2 E_{a} r_{A}\right)-\frac{1}{2} \bigg]
\end{equation}
where $\alpha$ is the fine-structure constant, $Z$ is the number of atom $A$, $g_{a \gamma \gamma}$ is the axion-photon coupling and $r_{A}$ is the radius of the atom. With this, we can estimate the sensitivity of our MeV axion from the axino decay in future neutrino experiments such as Hyper-Kamiokande. The target atom for the Hyper-Kamiokande is oxygen, whose atomic number is $Z=8$. There are $N_{O} \simeq 1.27 \times 10^{35}$ oxygen atoms for a fiducial mass of 3.8 Mt water~\cite{Abe:2018uyc}. By using the relation of the axion production rate $\Dot{N_{a}}$ and the differential axion flux $d\Phi_{a}/dE_{a}$~\cite{Payez:2014xsa}, we can obtain the expected number of events as
\begin{equation}
    N_{\rm event}=\int \frac{d\Phi_{a}}{dE_{a}} \sigma_{\rm el}\left(E_{a}\right) dE_{a} \times  \Delta t \times N_{O}
\end{equation}
where $\Delta t$ is the exposure time of Hyper-Kamiokande experiment and $r_{A}=2 \times 10^{-10}$m is the radius of the oxygen atom. 
\begin{figure}
\centering
\includegraphics[height=8cm,width=10cm]{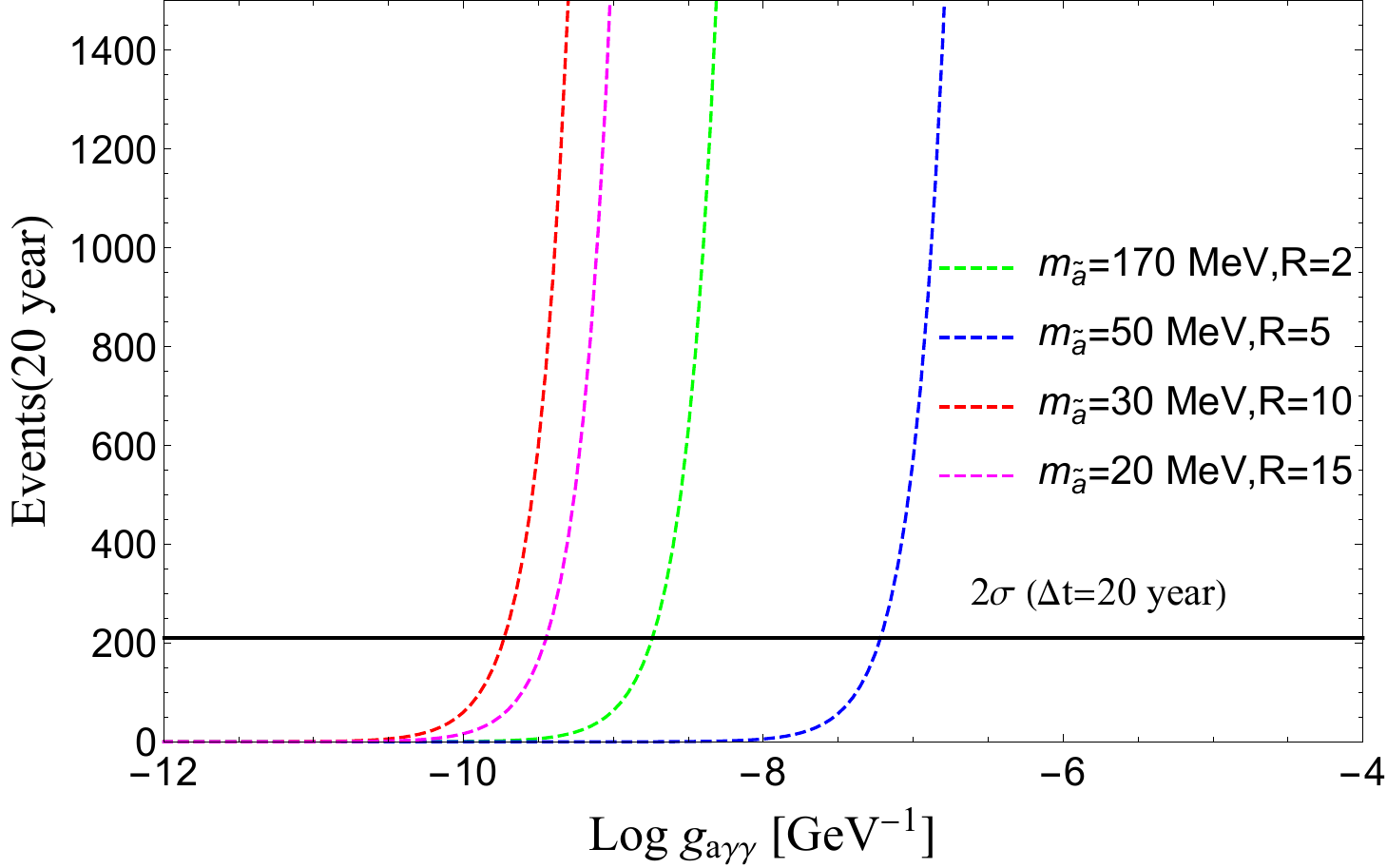}
\caption{Same benchmark points as Fig.~\ref{flux}, but for the number of annual events of the photon from the inverse Primakoff process $a+A \to \gamma+A$ as function of the axion-photon coupling $g_{a \gamma \gamma}$ in the Hyper-Kamiokande experiment. The black line denotes the $2\sigma$ exclusion limit.}
\label{events}
\end{figure}   

The main background is the steady diffuse supernova neutrino background (DSNB) arising from the distant core-collapse supernova. The observable events are mostly induced by the inverse beta decay ($\bar{\nu}_{e}+p \longrightarrow e^{+}+n$). Furthermore, the Kamiokande-II (Kam-II) experiment has observed the 12 events after the SN 1987A explosion. The steady detection rate of DSNB neutrino for Hyper-K experiment, related to the burst detection rate of SN 1987A neutrinos, can be simply estimated as~\cite{Beacom:2010kk}
\begin{equation}
    \Bigg[\frac{dN_{\nu}}{dt}\Bigg]_{DSNB}^{\rm HK} \sim \Bigg[\frac{dN_{\nu}}{dt}\Bigg]_{1987A}\Bigg[\frac{N_{SN} M_{\rm Kam}}{4 \pi D^2}\Bigg]_{1987A}^{-1} \Bigg[\frac{N_{SN} M_{\rm HK}}{4 \pi D^2}\Bigg]_{DSNB},
\end{equation}
where $M_{\rm Kam}$=2140 ton and $M_{\rm HK}$=3.8 Mton are the detector mass of Kam-II and Hyper-K experiment respectively. The detection rate in Kam-II was $[dN_{\nu}/dt]_{1987A} \sim 1$ s$^{-1}$ during the SN 1987A explosion. The distance $D$ of the SN 1987A was 0.050 Mpc, while $D= c/H_{0} \sim 4000$ Mpc for a typical supernova contribution to the DSNB. Besides, $N_{SN}=100$ for the DSNB whereas $N_{SN}=1$ for SN 1987A. Therefore, the time-averaged DSNB detection rate in Hyper-K is simply calculated as
\begin{equation}
    \Bigg[\frac{dN_{\nu}}{dt}\Bigg]_{DSNB}^{\rm HK} \sim (1 s^{-1}) \times 100 \times 10^{-10} \times 1800 \sim 568 \ \rm{year}^{-1}.
\end{equation}
The estimated backgrounds with exposure time 20 year for Hyper-K experiment are $\sim$11000. In Fig.~\ref{events}, we show the signal events as the function of axion-photon coupling $g_{a\gamma\gamma}$. The four benchmark points are the same as those in  Fig.~\ref{flux}. The black line denotes the $2\sigma$ bound on the coupling $g_{a\gamma\gamma}$. We can see that $g_{a \gamma\gamma}$ can be excluded down to $2 \times 10^{-10}$ GeV$^{-1}$ for our maximal flux case (red dashed line). Although such a limit is weaker than the constraint on low energy solar ALP from the CAST, $g_{a \gamma \gamma}<0.66 \times 10^{-10}$ GeV$^{-1}$~\cite{Anastassopoulos:2017ftl}, it may provide an independent way to hunt for the energetic axion ($E \sim {\cal O}({\rm MeV})$), which needs a more detailed study on the search strategy of this process.

\section{Conclusion}   
\label{sec6}

In this paper, we have investigated the dark sector of the supersymmetric DFSZ model, where the gravitino/axino is the late decaying DM. Owing to the saxion decay, the gravitino/axino problem that its relic density easily overclose the universe can be elegantly solved by the dilution effect. On the other hand, the early axion produced by the saxion decay as dark radiation contributes to the extra effective number of neutrino species $\Delta N_{\rm eff}$. However, due to the cosmological constraints on $\Delta N_{\rm eff}$ by CMB observations, increasing simply $\Delta N_{\rm eff}$ only alleviates the Hubble tension to $3 \sigma$. On the other, we find that a MeV axion emitted from the late decaying DM decay in the late universe can produce the sizable events of the inverse Primakoff scattering in the neutrino experiment like Hyper-Kamiokande via the inverse Primakoff scattering. We estimate the exclusion limit $g_{a\gamma\gamma} < 2 \times 10^{-10}$ GeV$^{-1}$ for an exposure of 3.8 Mton over 20 years. A more delicate experimental analysis of this signal may provide a way to probe the energetic axion.

\section{Acknowledgements}

We appreciate Pierluca Carenza for his help in the supernova neutrino background analysis. BZ is supported by the National Natural Science Foundation of China (NNSFC) under grant 
No. 11805161, by  Natural Science Foundation of Shandong Province under the grants ZR2018QA007, by the Basic Science Research Program through the National Research Foundation of Korea (NRF) funded by the Ministry of Education, Science and Technology (NRF-2019R1A2C2003738), and by the Korea Research Fellowship Program through the NRF funded by the Ministry of Science and ICT (2019H1D3A1A01070937).
\bibliography{refs}

\end{document}